\documentclass[acp, manuscript]{copernicus}

\usepackage{tikz}
\usetikzlibrary{patterns}
\usetikzlibrary{shapes}
\usetikzlibrary{decorations.text}
\usetikzlibrary{shapes.multipart}
\usetikzlibrary{arrows,shapes.geometric,positioning}
\usetikzlibrary{shapes,positioning,decorations.pathmorphing}
\usetikzlibrary{calc}
\usepackage{booktabs}
\usetikzlibrary{backgrounds}
\usepackage{varwidth}
\usetikzlibrary{fit}
\usepackage{xltabular}

\usepackage{bm,bbm}
\usepackage{amsmath,amsfonts,amssymb,mathrsfs}
\usepackage{latexsym,euscript,textcomp}
\usepackage{color,graphics,epsf,dcolumn}
\usepackage{epstopdf,ulem}
\usepackage{blindtext}
\usepackage{threeparttable}

\usepackage{enumitem}

\usepackage{orcidlink}

\newcommand\mc[1]{\multicolumn{1}{c}{\text{#1}}}
\newcommand\ml[1]{\multicolumn{1}{l}{\text{#1}}}

\newcommand{\overbar}[1]{\mkern 1.5mu\overline{\mkern-2.0mu#1\mkern-1.0mu}\mkern 1.5mu}

\newcommand{\beq}{\begin{equation}}
\newcommand{\eeq}{\end{equation}}
\newcommand{\pt}{\partial}
\allowdisplaybreaks

\begin{document}
\nolinenumbers

\title{The role of ecosystem transpiration in creating alternate moisture regimes by influencing atmospheric moisture convergence}


\Author[1,2]{Anastassia M.}{Makarieva\orcidlink{0000-0002-8598-5851}}
\Author[2]{Andrei V.}{Nefiodov\orcidlink{0000-0003-3101-4790}}
\Author[3]{Antonio Donato}{Nobre}
\Author[4]{Mara}{Baudena\orcidlink{0000-0002-6873-6466}}
\Author[5]{Ugo}{Bardi}
\Author[6,7,8]{Douglas}{Sheil\orcidlink{0000-0002-1166-6591}}
\Author[9]{Scott R.}{Saleska\orcidlink{0000-0002-4974-3628}}
\Author[10]{Ruben D.}{Molina\orcidlink{0000-0003-1548-2578}}
\Author[11]{Anja}{Rammig\orcidlink{0000-0001-5425-8718}}

\affil[1]{Institute for Advanced Study, Technical University of Munich, Lichtenbergstrasse 2~a, 85748 Garching, Germany}
\affil[2]{Theoretical Physics Division, Petersburg Nuclear Physics Institute, 188300 Gatchina, St.~Petersburg, Russia}
\affil[3]{Centro de Ci\^{e}ncia do Sistema Terrestre INPE, S\~{a}o Jos\'{e} dos Campos, 12227-010 S\~{a}o Paulo, Brazil}
\affil[4]{National Research Council of Italy, Institute of Atmospheric Sciences and Climate (CNR-ISAC), Torino, Italy}
\affil[5]{Department of Chemistry, University of Florence, Italy}
\affil[6]{Forest Ecology and Forest Management Group, Wageningen University \& Research, PO Box 47, 6700 AA, Wageningen, The Netherlands}
\affil[7]{Center for International Forestry Research (CIFOR), Kota Bogor, Jawa Barat, 16115, Indonesia}
\affil[8]{Faculty of Environmental Sciences and Natural Resource Management, Norwegian University of Life Sciences, \AA s, Norway}
\affil[9]{Department of Ecology and Evolutionary Biology, University of Arizona, Tucson, 85721, Arizona, USA}
\affil[10]{Escuela Ambiental, Facultad de Ingenier\'{i}a, Universidad de Antioquia, Medell\'{i}n, Colombia}
\affil[11]{Technical University of Munich, School of Life Sciences, Hans-Carl-von-Carlowitz-Platz 2, 
85354 Freising, Germany}


\runningtitle{Ecosystem transpiration and atmospheric moisture convergence}

\runningauthor{Makarieva et al.}

\correspondence{A. D. Nobre (anobre27@gmail.com), A. M. Makarieva (ammakarieva@gmail.com)}

\received{}
\pubdiscuss{} 
\revised{}
\accepted{}
\published{}


\firstpage{1}

\maketitle

\begin{abstract}
The terrestrial water cycle links the soil and atmosphere moisture reservoirs through four fluxes: precipitation, evaporation, runoff, and atmospheric moisture convergence (net import of water vapor to balance runoff). Each of these processes is essential for sustaining human and ecosystem well-being. Predicting how the water cycle responds to changes in vegetation cover remains a challenge. Recently, changes in plant transpiration across the Amazon basin were shown to be associated disproportionately with changes in rainfall, suggesting that even small declines in transpiration (e.g., from deforestation) would lead to much larger declines in rainfall. Here, constraining these findings by the law of mass conservation, we show that in a sufficiently wet atmosphere, forest transpiration can control atmospheric moisture convergence such that increased transpiration enhances atmospheric moisture import and resulting water yield. Conversely, in a sufficiently dry atmosphere increased transpiration reduces atmospheric moisture convergence and water yield. This previously unrecognized dichotomy can explain the otherwise mixed observations of how water yield responds to re-greening, as we illustrate with examples from China{\textquoteright}s Loess Plateau. Our analysis indicates that any additional precipitation recycling due to additional vegetation increases precipitation but decreases local water yield and steady-state runoff.  Therefore, in the drier regions/periods and early stages of ecological restoration, the role of vegetation can be confined to precipitation recycling, while once a wetter stage is achieved, additional vegetation enhances atmospheric moisture convergence and water yield. Recent analyses indicate that the latter regime dominates the global response of the terrestrial water cycle to re-greening. Evaluating the transition between regimes, and recognizing the potential of vegetation for enhancing moisture convergence, are crucial for characterizing the consequences of deforestation as well as for motivating and guiding ecological restoration.
\end{abstract}

\introduction[\color{black}Introduction]  

\label{intro}

\begin{figure*}[!b]
\begin{minipage}[p]{1\textwidth}
\centering\includegraphics[width=0.9\textwidth,angle=0,clip]{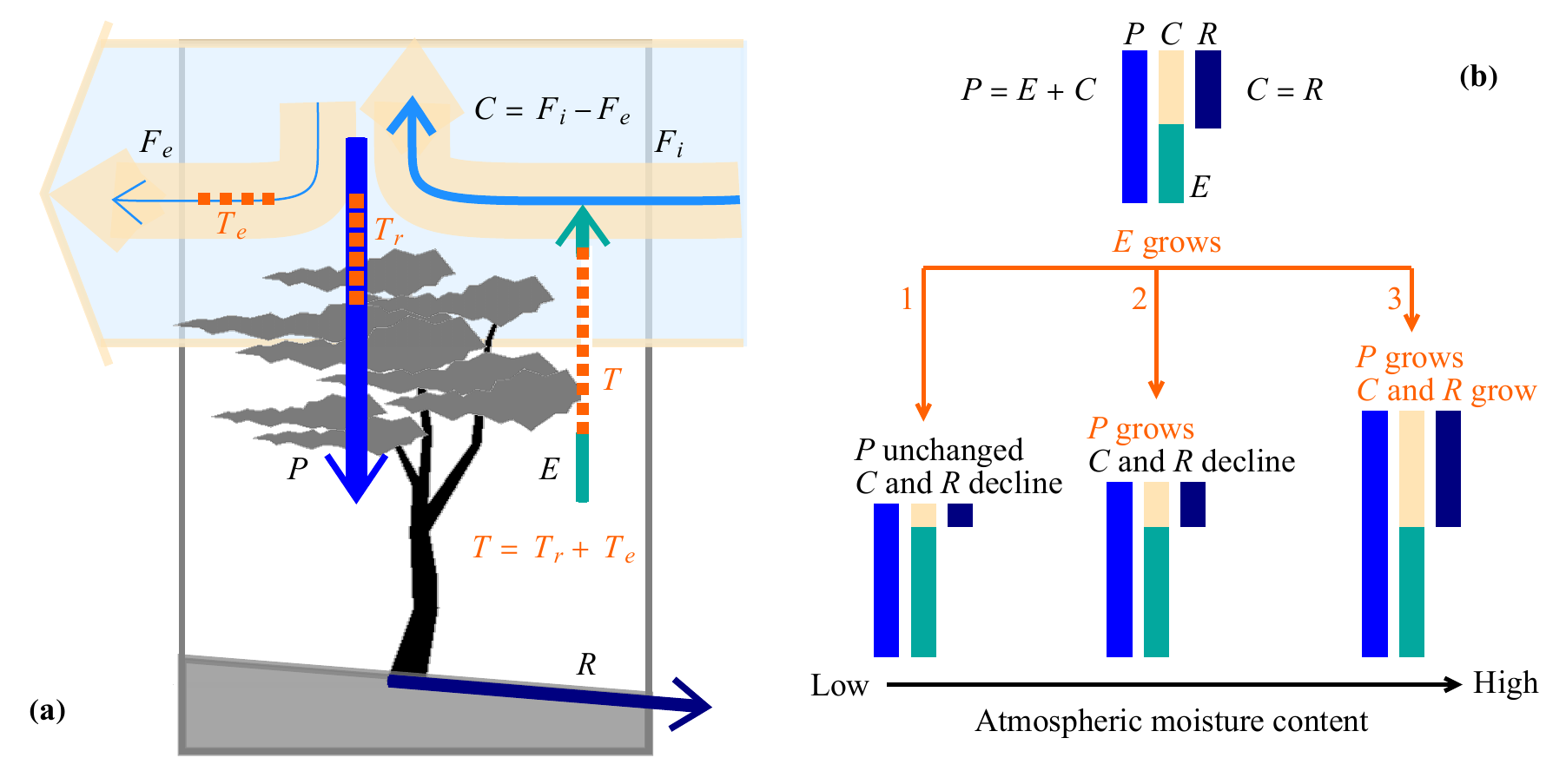}
\end{minipage}
\caption{
Steady-state moisture balance (a) and its possible changes with increasing evapotranspiration (b). In (a), the steady-state water balance, Eq.~\eqref{wbp}, is shown for a forest location that receives moisture solely from the ocean \citep[thin blue arrows indicate gross moisture import $F_{i}$ and  export $F_{e}$, cf.][their Fig.~12]{eltahir94}. Thick yellowish arrows indicate the ascending and descending air motions that generate precipitation $P$ and are responsible for local moisture convergence $C \ge 0$ (and steady-state runoff $R = C$). Orange dashing indicates fluxes of transpiration $T$, recycled transpired moisture $T_r$ and exported transpired moisture $T_e$.  Note that since $T \ge T_r$, increasing precipitation by $T_r$ (moisture recycling) reduces steady-state runoff $R$ by $T_e$. In (b), three possible cases are schematically shown of how the moisture balance can change in response to increasing evapotranspiration $E$ (the initial state is shown on top of the graph). Case 1 with constant (or declining) precipitation $P$ and declining runoff $R$ corresponds to the precipitation recycling approach on a local scale and the {\textquotedblleft}demand-side{\textquotedblright} argument of \citet{ellison2012}, case 2 corresponds to the precipitation recycling approach on a larger scale and the {\textquotedblleft}supply-side{\textquotedblright} argument of \citet{ellison2012}, and case 3 corresponds to moisture convergence controlled by plant transpiration.
}
\label{fig1}
\end{figure*}

Land loses water through runoff to oceans. The terrestrial water cycle depends
on atmospheric moisture convergence: the water that arrives via the atmosphere to compensate for runoff. This water is delivered to land by the wind in the form of water vapor.
When moist air rises over the land, it cools, and water vapor condenses and precipitates. If the air does not rise and cool, the airflow carries the water vapor further downwind. Atmospheric moisture convergence is the term used to express difference between gross import and export of water vapor in the region of interest. 
When the atmospheric and groundwater storages are steady-state,
atmospheric moisture convergence equals runoff (Fig.~\ref{fig1}a). 

Forests play a major role in the water cycle, which deforestation and reforestation can potentially alter in multiple ways. 
Atmospheric moisture convergence $C$ and evapotranspiration $E$ add moisture to the atmosphere, while precipitation $P$
removes it. The steady-state water budget for the atmosphere is
\beq\label{wbp}
P = E + C.
\eeq
If precipitation is constant, then increased evapotranspiration due to reforestation will diminish atmospheric moisture convergence and runoff
(Fig.~\ref{fig1}b, case $1$). This can potentially compromise freshwater availability, irrigation, navigation, and hydropower.
If atmospheric moisture convergence is constant, increased evapotranspiration will elevate rainfall.
This, conversely, will benefit freshwater availability and agriculture.
\citet{ellison2012} grouped these contrasting responses into {\textquotedblleft}demand-side{\textquotedblright} versus {\textquotedblleft}supply-side{\textquotedblright}, respectively.
Published syntheses reveal diverse hydrological responses to land cover change that depend on the spatiotemporal scales and locations \citep[e.g.,][and references therein]{lawrence15,tewierik2021,posada-marin2022}. For example, deforestation in Brazil can increase local rainfall but leads to declines on a larger scale \citep{leitefilho21}. 
During the massive afforestation on the Loess Plateau in China, atmospheric moisture convergence initially diminished but later began to increase \citep{zhang2022}.

Compared to this diversity, the prevailing approach to studying how changes in evapotranspiration impact the water cycle has been relatively uniform,
based on the concept of precipitation recycling \citep[e.g.,][]{salati79,eltahir94,sav95,ent10,ke12,ellison2012,zemp17b,zemp17a,staal2018,wang-erlandsson2018,hoekvandijke2022}.
Evapotranspiration returns water to the atmosphere. Some of this
evapotranspired water vapor may condense and precipitate again, which can be seen as precipitation recycling in the considered area.
The remaining vapor will be blown away by the wind.
Recycled precipitation is, therefore, always smaller than, or equal to, evapotranspiration \citep[see Fig.~\ref{fig1}a and discussion by][]{eltahir94}.

The conventional precipitation recycling approach assumes that the air circulation (i.e., wind speed and direction) does not change in response to land cover change.
Water vapor is viewed as a tracer that merely follows the airflow.
This implies that if evapotranspiration diminishes/increases by a certain amount, precipitation is diminished/increased by a smaller amount of recycled precipitation.
Consequently, this approach invariably predicts that atmospheric moisture convergence (and steady-state runoff) 
will decline with increasing evapotranspiration (and rise with declining evapotranspiration upon deforestation) (Fig.~\ref{fig1}, case $2$).
(As an example, for a large-scale continental reforestation, the precipitation recycling approach predicts a considerable decline in global runoff as a result of increased evapotranspiration \citep{hoekvandijke2022}, but see our discussion in Section~\ref{sec4.1} below.) 
This implies that competition for water resources between added vegetation and humans
is unavoidable \citep{ricciardi2022}. This implication has particular relevance in China, where the hydrological impacts of large-scale re-greening programs are
actively debated \citep{feng2016,zheng2021,zhang2022}.
Conversely, reduced evapotranspiration associated with elevated concentrations of carbon dioxide was expected to increase moisture convergence and runoff
\citep[e.g.,][]{gedney06}. 

Besides precipitation recycling and consideration of water vapor as a tracer, it has been recognized that change in evapotranspiration can cause changes in atmospheric moisture convergence
by modifying the surface temperature and atmospheric temperature profile \citep[e.g.,][]{pu14,eiras-barca20}. However,
the direction of such changes has long been a matter of controversy, as illustrated, for example, by an early discussion between \citet{ripley76} and \citet{charney76} (see also \citet{makarieva22} for an overview) and more recently by modelling studies revealing diverse responses in different regions \citep[e.g.,][]{kooperman18}.
Accordingly, increase (reduction) of moisture convergence with reduced (increased) evapotranspiration has been acknowledged as the first-order effect in the water cycle response to  
changes in evapotranspiration \citep[][and references therein]{kooperman18,fowler19}.

Our goal in the present paper is to develop a more comprehensive
conceptual approach taking into account that vegetation can alter atmospheric dynamics and hence the water cycle
via its impact on atmospheric moisture content -- such that both precipitation and moisture convergence will grow in response to increasing evapotranspiration (Fig.~\ref{fig1}b, case $3$).   Our key point is that a moist atmosphere close to saturation displays different dynamics compared to a dry, unsaturated atmosphere. As increased evapotranspiration
leads to higher atmospheric moisture content, this can elicit dynamics with both precipitation and moisture convergence enhanced.
From this more comprehensive perspective, the strategic question for any large-scale ecological restoration effort will not be
{\textquotedblleft}When/where should we stop/avoid ecological restoration to not run into water limitation?{\textquotedblright} \citep[cf.][]{feng2016,ricciardi2022}. It will be {\textquotedblleft}What conditions are necessary for restoration to maximally enhance the water cycle (both precipitation and runoff)? How can we plan restoration to achieve these conditions?{\textquotedblright}. 

\begin{figure*}[!tb]
\begin{minipage}[p]{1\textwidth}
\centering\includegraphics[width=0.85\textwidth,angle=0,clip]{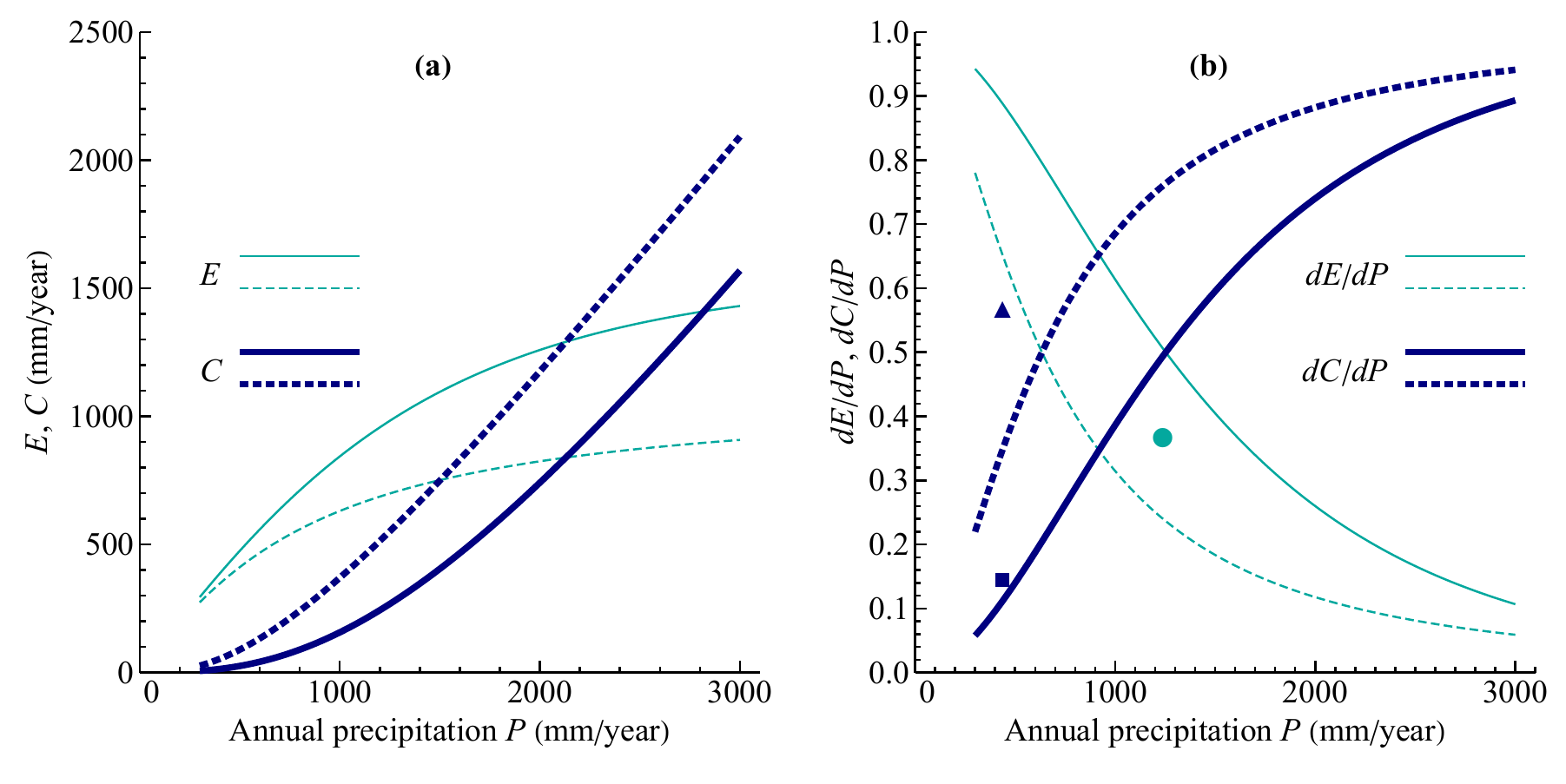}
\end{minipage}
\caption{
The dependence of annual evapotranspiration $E$ (a) and $dE/dP$ (b) (thin green lines) and moisture convergence $C = P - E$ (a) and $dC/dP$ (b) (thick blue lines) on annual precipitation $P$ in forest (solid lines) and grass (dashed lines) ecosystems according to \citet{zhang2001}, see Eq.~\eqref{zh}.
The green circle indicates $dE/dP$ in the Southern Amazon in the late dry season (see Section~\ref{SA}).
The blue triangle and square indicate, respectively, estimated $dC/dP$ and measured $dR/dP$ on the Loess Plateau in $1999$--$2015$ according to \citet{zhang2022} (see Section~\ref{LP}).
}
\label{fig2}
\end{figure*}

There are multiple independent lines of evidence testifying that a simultaneous enhancement of all moisture flows by vegetation is plausible. 
Generally, ecosystems with higher evapotranspiration tend to have higher atmospheric moisture convergence and runoff. In Fig.~\ref{fig2}, the dependence of annual evapotranspiration on annual precipitation is shown for forests and grasslands as established by \citet{zhang2001} based on catchment water balance \citep[see also][]{teuling2018}. From this dependence and the steady-state water budget, one can see that atmospheric moisture convergence and runoff increase with evapotranspiration. In the Amazon forest, increased forest transpiration at the end of the dry season leads to the accumulation of atmospheric moisture. This moistening facilitates convection, which, in turn, boosts atmospheric moisture convergence \citep{wright17}. 

Furthermore, previous studies established the link between atmospheric water content and convection and precipitation.  \citet{bretherton04} found that observations of tropical oceanic daily and monthly mean precipitation depend exponentially on atmospheric moisture content. Based on data for a tropical island, \citet{holloway10} showed that the rainfall probability rises sharply with increasing atmospheric moisture content \citep[see also][]{yano22}.  Using the radiosonde data for several meteostations in Brazil (Fig.~\ref{fig3}a) and the relationship
established by \citet{holloway10}, \citet{jhm14} found that, in the Amazon forest with its wet atmosphere, a small increment in 
the atmospheric moisture content leads to a several times{\textquoteright}  larger increment in rainfall probability than in the drier regions.  \citet{jhm14} suggested that, due to its moist atmosphere, local Amazonian vegetation can trigger or suppress convection and rainfall by changes in transpiration, emission of biogenic condensation nuclei, and other biotically mediated processes.
\citet{baudena21} used reanalyses data in a part of the Amazon basin to establish that hourly precipitation grows non-linearly with increasing atmospheric moisture content. 

Finally, \citet{mapes18} showed that the persistence of spatial domains with high atmospheric moisture and high precipitation rates in the tropics
testifies to their maintenance by atmospheric moisture convergence. Where atmospheric moisture content and precipitation exceed a certain threshold, such an area becomes
self-sustainable. It generates a sufficient inflow of moist air to keep the high moisture content and to feed the high precipitation that significantly exceeds
local evaporation. Therefore, all the three processes necessary 
for realizing case 3 in Fig.~\ref{fig1}b, have been independently reported on different spatial and temporal scales: 1) increased evapotranspiration leads to
increased atmospheric moisture content; 2) increased atmospheric moisture content leads to a sharp increase in precipitation; 3) high 
precipitation induces and/or sustains efficient atmospheric moisture convergence.
Here we combine and extend these analyses and conceptualize them into a new comprehensive approach to describing vegetation{\textquoteright}s impact 
on the water cycle. We shall see that this approach accommodates the precipitation recycling approach as a special case.

\section{Data and methods}

\subsection{Key variables and relationships}
\label{theor}

We quantify local atmospheric moisture content as column water vapor $W$ (mm):
\beq\label{W}
W \equiv \frac{1}{\rho_l}\int_0^{z_{\rm T}} \rho_v dz.
\eeq
Here $\rho_v$ is the density of water vapor and $\rho_l = 10^3$~kg/m$^{3}$ is the density of liquid water,
and the integration is performed over the entire atmospheric column from the Earth{\textquoteright}s surface to the top of the troposphere $z_{\rm T}$,
where water vapor density is negligible.
Condensed water makes a negligible contribution to the atmospheric moisture content \citep{jas13}.
Due to the $1/\rho_l$ factor, $W$ measures atmospheric water content in terms of equivalent water depth: $1$~mm  is equivalent to $1$~kg of water vapor per squared meter.
Local groundwater storage $G$ (mm) is defined similarly to $W$.
\beq\label{G}
G \equiv \frac{1}{\rho_l}\int_{z_{\rm G}}^{\rm 0} \rho_g dz,
\eeq
where $\rho_g$ is the mean water density in the ground (water mass per unit ground volume) and the absolute value of $z_{\rm G}$ is the depth
of the considered ground layer ($z_{\rm G} < 0$).

The mean value of $W$ for a region of area $S$ and atmospheric volume $V = z_{\rm T}S$ is
\beq\label{Wm}
W \equiv \frac{1}{\rho_lS}\int_{V} \rho_v dV.
\eeq

The atmospheric moisture convergence $C$ in the considered region is determined as follows:
\beq\label{C}
C   \equiv -\frac{1}{\rho_lS} \int \limits_{V} \mathrm{div} (\rho_{v} \mathbf{u}) dV= -  \frac{1}{\rho_l S} \oint\limits_{\sigma} \rho_{v} (\mathbf{u}\cdot \mathbf{n}) d\sigma \equiv F_i - F_e, 
\eeq
where  $\mathbf{u}$ is the vector of air velocity. The closed surface $\sigma$ encloses volume $V$.  The unit  normal vector $\mathbf{n}$ is directed outward. Flux $F_i$ corresponds to the case when the normal component of air velocity $u_n = (\mathbf{u}\cdot \mathbf{n})<0$, i.e., the water vapor flows into the region.  
Flux $F_e$ corresponds to the case when $u_n  > 0$, i.e., the water vapor flows out of the region. 
Since $\rho_v \simeq 0$  at $z = z_{\rm T}$ and $u_n = 0$ at $z = 0$, atmospheric moisture convergence
$C$ describes the net flux of water vapor through the lateral surface of the
considered atmospheric volume (Fig.~\ref{fig1}a).  Note that the precipitation recycling approach assumes that the air velocity does not change upon reforestation/deforestation, but only $\rho_v$ does. 

The difference between evaporation and precipitation is
\beq\label{dpe}
E -P   \equiv \frac{1}{\rho_l S} \int \limits_{V} \dot{\rho}_{v}  dV ,
\eeq
where $\dot{\rho}_{v}$ is the mass source/sink of water vapor, i.e., evaporation if $\dot{\rho}_{v} > 0$ and condensation if $\dot{\rho}_{v} < 0$. The quantities $C$,  $E$, and $P$ are measured in units of length per unit of time (for example, mm/day). 
Local values of $C$ and $E-P$ are obtained by substituting the volume integration in Eqs.~\eqref{C} and \eqref{dpe} by one-dimentional integration
over altitude $z$, cf. Eqs.~\eqref{W} and \eqref{Wm}.

The mass balance for the atmospheric moisture and groundwater can be written as follows:
\begin{eqnarray}\label{wb1}
\frac{dW}{dt} = C + E - P,\\ \label{wb2}
\frac{dG}{dt} = P - E - R.
\end{eqnarray}
Here $P$ is precipitation, $E$ is evapotranspiration, and $R$ (mm/day) is runoff.
These mass balance equations apply both locally and on a regional scale. 

The steady-state mass balance for the atmosphere, Eq.~\eqref{wbp}, is
valid when the rate of change in the atmospheric moisture content is low compared to precipitation ($dW/dt \ll  P$).
This pertains to time scales $\tau \gtrsim \tau_W$, where $\tau_W \equiv W/P$ is the time scale of atmospheric moisture 
turnover via precipitation; $\tau_W$ is about ten days on a global average \citep{laederach16}. 
The groundwater storage $G$ is several orders of magnitude larger than $W$. The so-called modern 
groundwater (recycled within the last $50$ years) constitutes about $G = 3$~m on a global average, and its mean turnover time via global
runoff $R = 0.5$~m/year is $\tau_G \equiv G/R = 6$~years $ \gg \tau_W$ \citep{lv79,gleeson2016}.
On the time scales when both atmosphere and soil moisture are steady-state, runoff equals moisture convergence, $C = R$ (Fig.~\ref{fig1}). When
the groundwater storage is not steady, then $C = P - E = R + dG/dt$, whereby $R+dG/dt$ is referred to as the water yield \citep{zhang2022}
or water availability \citep{cui22}.

\begin{figure*}[!tb]
\begin{minipage}[p]{1\textwidth}
\centering\includegraphics[width=0.8\textwidth,angle=0,clip]{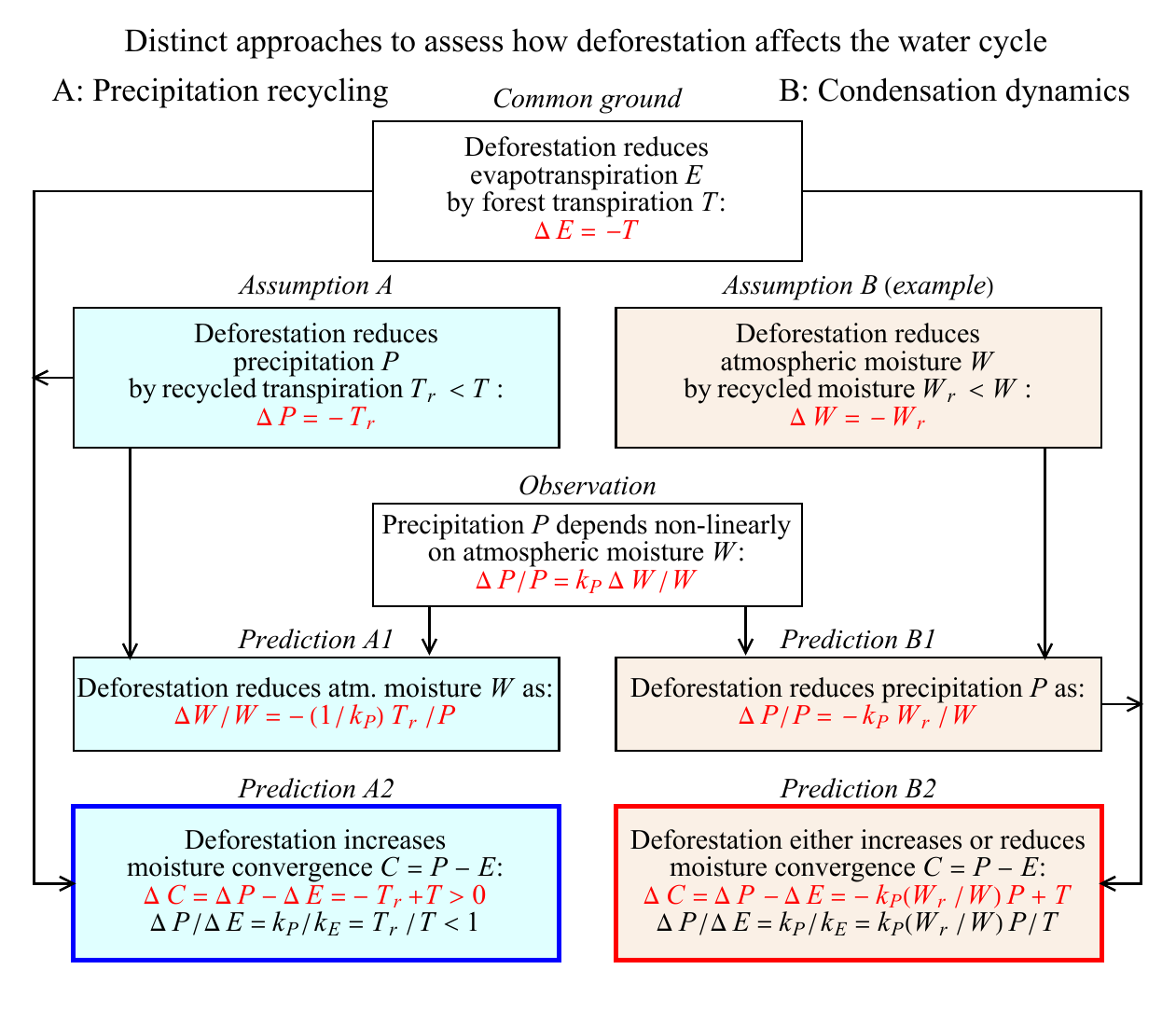}
\end{minipage}
\caption{
Precipitation recycling (A) versus condensation dynamics (B) approaches, their key assumptions and predictions concerning deforestation. Approach A focuses on flows (Assumption A),
Approach B focuses on atmospheric moisture content (Assumption B). Predictions of the two approaches with respect to atmospheric moisture
convergence coincide in a special case when $T_r/P = k_P W_r/W$. To describe reforestation, the minus signs on the right-hand side of all the equations (except the lowest ones) should be replaced by the plus signs, and vice versa. For reforestation ($\Delta E > 0$), the precipitation recycling approach predicts $\Delta C < 0$. 
}
\label{conc}
\end{figure*}

Dynamics is introduced by relating the atmospheric moisture content $W$ (mm) to precipitation $P$ (mm/day).
Since there is no fundamental constant to relate these two magnitudes of different dimensions, the dependence between them
is formulated in terms of their dimensionless relative increments $dW/W$ and $dP/P$ \citep[][Chapter~2.6]{vg95}: 
\beq\label{P}
\frac{dP}{P} = k_P(W)\frac{dW}{W}. 
\eeq
In biology, this type of relationships is widely used and known as allometric.
Here dimensionless $k_P(W)$ is the slope of the $P$-$W$ dependence on a log-log scale. 
Observations show that $k_P(W) \ge 0$, i.e., precipitation rises with increasing atmospheric moisture content (see the next section).

The proposed control of atmospheric moisture convergence by evapotranspiration can occur as follows.
At a certain point of time $t = t_0$ evapotranspiration begins to rise, $dE/dt > 0$.
According to the mass balance equation \eqref{wb1}, 
an increase in $E$ leads to an increase in the rate of change of the atmospheric moisture content,
$d^2W/dt^2 = dE/dt > 0$. {\it If} at the initial moment of time $dW/dt = 0$, this rate becomes positive as time goes,
$dW/dt = (dW/dt)|_{t =t_0} + E(t) - E(t_0) = \Delta E(t) > 0$. Consequently, $W$ itself also grows, $\Delta W(t) > 0$.

Without losing generality, for the relationship between evapotranspiration and moisture content we can write
\beq\label{E}
\frac{dE}{P} = k_E(W)\frac{dW}{W}. 
\eeq

As the moisture content increases, so, according to Eq.~\eqref{P}, does precipitation. The physical mechanisms that
relate increased precipitation to increased atmospheric moisture convergence can cause the moisture convergence to rise as well.
We do not consider these mechanisms here. According to the mass balance equation, an increase in precipitation leads to a decline
in the rate of change of the atmospheric moisture content: other things being equal, the increase of $dW/dt$ slows down when $dP/dt > 0$. When the accumulation of atmospheric moisture ceases, a new steady state with $dW/dt = 0$ is reached with a higher $W$, $E$, and $P$.

When the ecosystem goes from one steady state with evapotranspiration $E$ to another steady state with altered evapotranspiration $E + \Delta E$, from the mass balance equation \eqref{wbp} we have
\beq\label{wbc}
\Delta C = \Delta P - \Delta E = P \left(k_P - k_E\right) \frac{\Delta W}{W} = \left(k_P- k_E \right)\frac{\Delta E}{k_E}.
\eeq
Moisture convergence will rise if precipitation increases more rapidly than evapotranspiration: $\Delta C > 0$ if $\Delta P > \Delta E$.
More specifically, it will rise with increasing evapotranspiration $\Delta E > 0$ if $k_P > k_E > 0$.

We will term the approach based on Eq.~\eqref{wbc} and the consideration of atmospheric moisture content, the {\it condensation dynamics} approach -- as we will see, there are different states of the water cycle depending on whether the atmosphere is moist enough to permit condensation. Equation~\eqref{wbc} indicates that,  to know how moisture convergence changes, distinct constraints on the behavior of evapotranspiration and precipitation with changing $W$, i.e.,  constraints determining $k_E$ and $k_P$, are required. In this comprehensive approach, deforestation or reforestation can either increase or decrease moisture convergence, depending on specific conditions. In contrast, the precipitation recycling approach uniquely predicts a decline in moisture convergence upon reforestation. Figure \ref{conc} summarizes the conceptual difference between the two approaches.

\subsection{Data description}

\begin{figure*}[!tb]
\begin{minipage}[p]{1\textwidth}
\centering\includegraphics[width=0.45\textwidth,angle=0,clip]{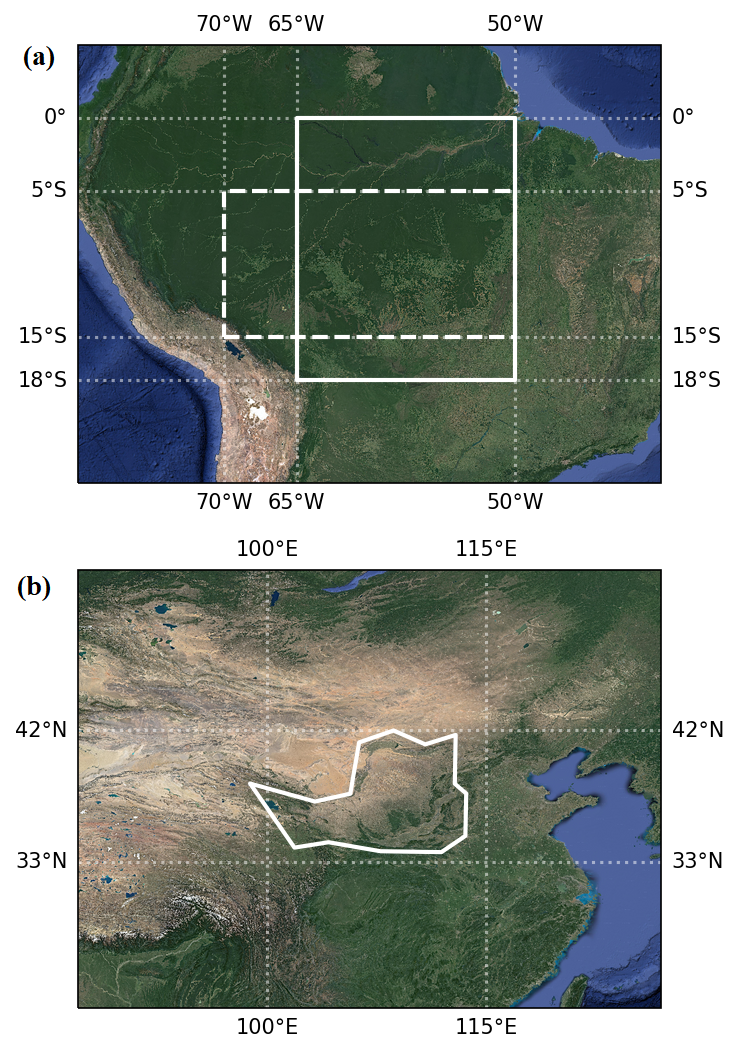}
\end{minipage}
\caption{
Regions from which the data were used: (a) the Amazon rainforest, solid indicates the region studied by \citet{baudena21} (hourly precipitation),
dashed rectangle -- the region studied by \citet{wright17} (five days{\textquoteright} precipitation); (b) the Loess Plateau in China with borders shown as in Fig.~4a of \citet{tian2022}
whose multi-year data on moisture convergence, precipitation, and evapotranspiration we used. Imagery \textcopyright 2022 TerraMetrics, Map data \textcopyright 2022 Google.
}
\label{figmap}
\end{figure*}

To study processes formalized by Eqs.~\eqref{C}--\eqref{wbc}, we used the data for the Amazon rainforest,
as the largest and best studied tropical forest region, and for the Loess Plateau in China, as the largest and best studied
long-term experiment on ecological restoration (Table~\ref{dat} and Fig.~\ref{figmap}).

\begin{table}[htbp]
    \caption{Summary of the data analyzed}\label{dat}
    \begin{threeparttable}
    \centering   
        \begin{tabular}{l | l  l l }
   \toprule                 
Process    &  Hypothetical deforestation  &    Onset of the wet season  &    Long-term  regreening  \\  
 Region    &  Amazon  &  Amazon & The Loess Plateau \\
              &      0--18$\degree$S, 50--65$\degree$W           &  5--15$\degree$S, 50--70$\degree$W               &  China \\
 Study period    &  2003--2014   &   2005--2011   &  1982--2018  \\       
Temporal resolution            &  hourly   &   5 days    &  annual  \\      
 Key data sources   &  ERA5  &   ERA-Interim, TRMM, AIRS   &  MERRA-2 \\       
 Key references    &  \citet{baudena21}  &  \citet{wright17}   &  \citet{zheng2021};   \\    
  &    &&  \citet{zhang2022};     \\        
  &    &&    \citet{tian2022} \\   
 \bottomrule                           
     \end{tabular} 
\begin{tablenotes}[para,flushleft]
$\,$
\end{tablenotes}
\end{threeparttable}
\end{table}     

For the Amazon rainforest, we first established the dependence between precipitation and atmospheric moisture content, Eq.~\eqref{P}, using local hourly data.
We acquired ERA5 reanalysis data \citep{hersbach2018} for a large and relatively flat study area between 0--18$\degree$S and 50--65$\degree$W, following the methods of \citet{baudena21}.  Reanalysis data combine and refine/adjust observations with short-term model-derived simulations to interpolate values and ensure consistent and complete (gap-free) global coverage. The studied area includes the Amazon forest and a part of the transition zone to the Cerrado (i.e., a savanna), also encompassing agricultural areas. Hourly precipitation $P$ (mm/hour) and column water vapor $W$ (mm) were acquired for 2003--2014 at a $0.25\degree$ resolution (Copernicus Climate Change Service).

Similarly to \citet{baudena21}, we binned hourly precipitation data for every 1 mm of total column water vapor $W$ for the study area, and we calculated the average of the distribution of the precipitation for each bin. For each bin, we used in our analysis the average $P$ values only if there were more than $20,000$ points in the bin or about $0.005\%$ of the observations (see inset in Fig.~\ref{fig3}b), thus retaining $W$ values from $7$~mm to $67$~mm. Additional data on the statistics of precipitation values in each bin (percentiles and standard deviation)
are presented in the Supplementary Information (Table~S1).

To compare the obtained $P(W)$ dependence with the previous findings, in Fig.~\ref{fig3} we plotted the empirically fitted exponential dependence of daily $P_d$ (mm/day) and monthly $P_m$ (mm/month) oceanic precipitation on $W$ (mm) according to, respectively, Eqs.~(1) and (2) of \citet{bretherton04}:
\beq\label{br}
P_d(W) = \exp[0.22 (W - 43)], \qquad P_m(W) = \exp[0.16 (W - 38)].
\eeq
\citeauthor{bretherton04}{\textquoteright}s~\citeyearpar{bretherton04} Eqs.~(1) and (2) relate $P$ to the relative moisture content $W/W^*$, where $W^*$ is the saturated moisture content for a given atmospheric sounding. To obtain Eq.~\eqref{br} we used the characteristic mean $W^* = 72$~mm \citep[][p.~1521]{bretherton04}. This introduces an uncertainty associated with a possible dependence of $W$ on temperature.

We analyzed the dependence between precipitation and atmospheric moisture content on a seasonal scale using the data from Fig.~S7 of
\citet{wright17} for the Southern Amazon (Fig.~\ref{figmap}).
These data include evapotranspiration retrieved from ERA-Interim reanalysis \citep{dee2011}, precipitation retrieved from
TRMM 3B42 daily gridded precipitation product at $0.25\degree \times 0.25\degree$ resolution \citep{huffman2007}
and atmospheric moisture content retrieved from Version 6 of the AIRS Level 3 daily gridded product at $1\degree \times 1\degree$ resolution \citep{olsen2017,tian2017}.
All analyzed variables are area-weighted spatial averages for the region bounded by $5\degree$S to $15\degree$S and $50\degree$W to $70\degree$W. 
Variables are averaged over discrete five days{\textquoteright} periods (pentads) for years from $2005$ to $2011$. In each year, the pentads are aligned
relative to the onset of the wet season defined as the first pentad when the rain rate exceeded the climatological
mean and the rain rate in at least five of the eight preceding (subsequent) pentads was less (greater) than the climatological mean.

For the Loess Plateau, we used the data from Fig.~10 of \citet{tian2022}, which include
evapotranspiration, precipitation, atmospheric moisture content and atmospheric moisture fluxes $F_{i}$ and $F_e$
from MERRA-2 reanalysis (monthly data at $0.5\degree \times 0.625\degree$ resolution) for the years from $1982$ to $2018$ for the rainy season (from June to September).

Additionally, we used the data from the analyses of \citet{jhm14} and \citet{baker21a} for the Amazon and of \citet{zheng2021}, \citet{zhang2022}, and \citet{dong2019} for the Loess Plateau as indicated, respectively, in the legends to Fig.~\ref{fig3}, Fig.~S1, Fig.~\ref{fig6}, and Fig.~\ref{fig7}.

For investigation of the $E$-$W$ dependence, we used the empirical parameterization of \citet{zhang2001} (Fig.~\ref{fig2}):
\beq\label{zh}
E(P) = E_0 \frac{\xi (\xi+a)}{\xi^2 + \xi +a} , \quad \xi \equiv \frac{P}{E_0} , 
\eeq
where for forests $E_0 = 1410$~mm/year and $a = 2$, while for grasses $E_0 = 1100$~mm/year and $a = 0.5$.

\section{\color{black}Results}

\subsection{Dependence of precipitation on atmospheric moisture content}

\begin{figure*}[!tb]
\begin{minipage}[p]{1\textwidth}
\centering\includegraphics[width=0.85\textwidth,angle=0,clip]{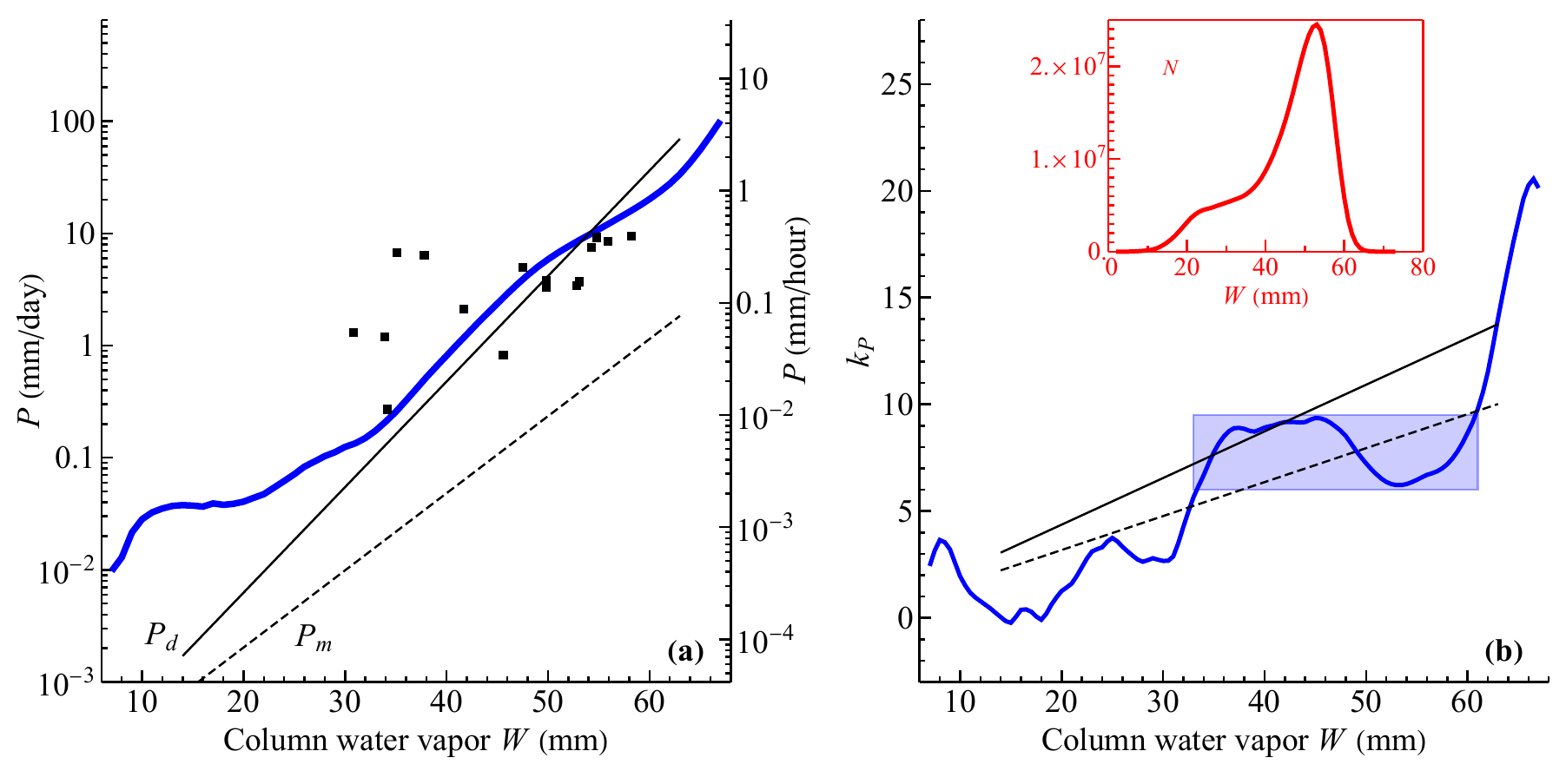}
\end{minipage}
\caption{
Precipitation $P(W)$ (a) and $k_P(W)$ (b) for the data of \citet{baudena21} (thick blue lines) and \citet{bretherton04} (thin black lines, daily (solid) and monthly (dashed), Eq.~\eqref{br}). Squares in (a) represent daily averaged $P$ and $W$ for dry and wet seasons at several meteostations in Brazil \citep[][Table~2, columns 4 and 11]{jhm14}. The inset in (b) shows the number of data points $N(W)$ for each 1 mm bin  \citep{baudena21}. The values of $k_P$ are obtained from Eq.~\eqref{P} using the $P(W)$ curves shown in (a). The blue rectangle indicates the interval of relatively constant $k_P$. 
}
\label{fig3}
\end{figure*}

The established relationship \eqref{P} between hourly precipitation $P$ and atmospheric moisture content $W$ in the Amazon
is shown in Fig.~\ref{fig3}. It displays three distinct features. At low moisture content, precipitation is practically constant and independent of $W$:
$k_P \sim 0$ for $10~{\rm mm}< W < 20$~mm.  At an intermediate moisture content of $30~{\rm mm}< W < 60$~mm (corresponding to most observations, 
see the inset in Fig.~\ref{fig3}b), precipitation increases several times faster than moisture content: $6 \lesssim k_P \lesssim 9$. At the highest moisture content $W \ge 60$~mm, there is an even more abrupt
increase in precipitation, corresponding to a sharp rise in $k_P$ up to $k_P > 20$. 

In comparison, the exponential dependence used to approximate the relationship between daily and monthly precipitation and atmospheric moisture
content over the ocean \citep{bretherton04} corresponds to a linear increase of $k_P$ over the entire range of moisture content values (Fig.~\ref{fig3}b).
The exponential model does not permit the identification of moisture content intervals with distinct behavior of $k_P$.

Let us discuss the physical meaning of these three features.
At the lowest moisture content values, the dry atmosphere is far from saturation. Under such conditions, water vapor behaves merely as a tracer
that cannot perturb but passively follows atmospheric dynamics. Accordingly, precipitation (which reflects these dynamics) is independent of atmospheric moisture content.
\citet{jhm14} discussed evidence for how, during the drier periods and in drier locations in Brazil, precipitation is, rather, determined by non-local weather systems.

\begin{figure*}[!tb]
\begin{minipage}[p]{1\textwidth}
\centering\includegraphics[width=0.5\textwidth,angle=0,clip]{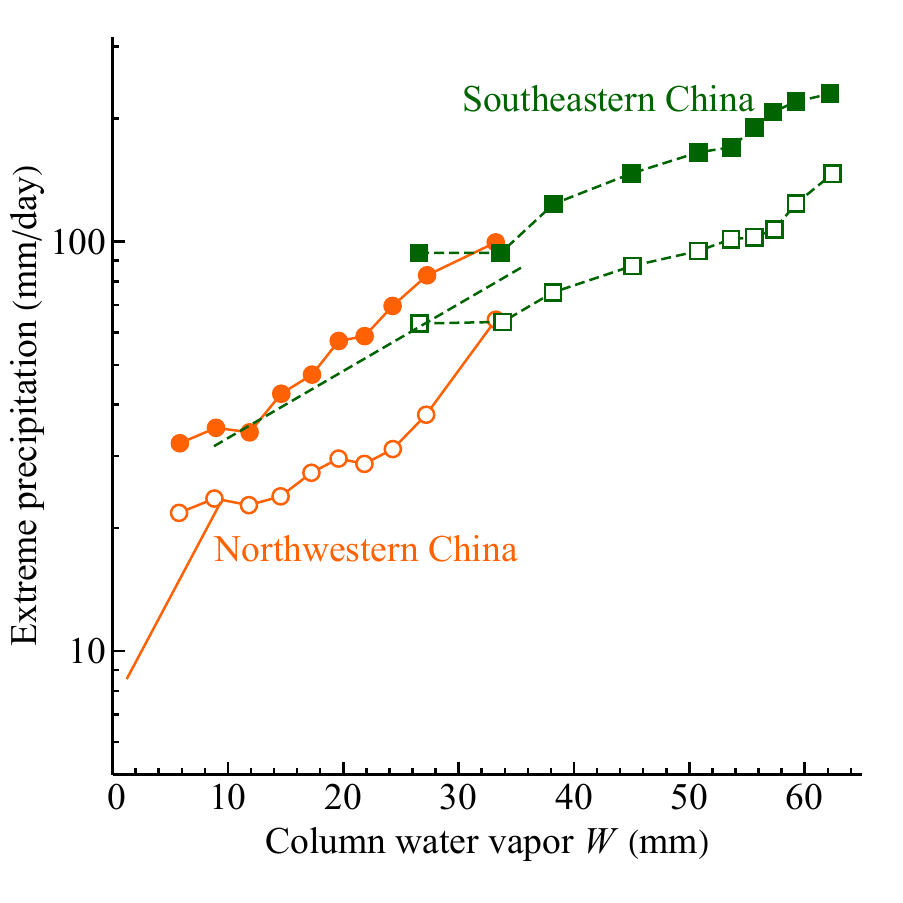}
\end{minipage}
\caption{
Extreme precipitation in northwestern (orange solid lines and circles) and southeastern (green dashed lines and squares) China, 
data taken from Fig.~6 of \citet{dong2019}. Closed and open symbols show the $99.9$  and $99$ percentiles for summer precipitation. The $99.9$ percentiles for winter precipitation are shown as straight lines using the exponential fit obtained by \citet{dong2019}.
}
\label{fig7}
\end{figure*}

As low $k_P \sim 0$ is associated with unsaturated conditions, 
this feature in seasonal climates should become pronounced only after temperature and relative humidity 
considerations are taken into account \citep[cf. discussion by][]{bretherton04}.
For example, according to the data of \citet[][]{dong2019}, extreme summer rainfall in various regions of China 
is relatively constant at the lower end of observed moisture content values: 
$5~{\rm mm}\lesssim W \lesssim 15$~mm for the colder northwestern regions and $25~{\rm mm}\lesssim W \lesssim 35$~mm for the warmer
southeastern regions (Fig.~\ref{fig7}).

In winter, due to the lower temperature, the same moisture content intervals correspond to a higher relative humidity and the atmospheric column is closer to saturation. Accordingly, for winter rainfall there is no leveling off at these values (Fig.~\ref{fig7}). Merging the winter and summer rainfall leads to the disappearance of this feature. 
 It is noteworthy that the limited response of extreme summer rainfall to
changes in moisture content is registered in both northwestern and southeastern regions of China that differ greatly in their annual precipitation and atmospheric moisture contents \citep{dong2019}.
(\citet{dong2019} mention that the linear slopes in the $P$-$W$ dependencies are {\textquotedblleft}less than one{\textquotedblright} which, according to them, is in agreement with {\textquotedblleft}indications that lower tropospheric moisture content increases faster than rainfall{\textquotedblright}. However, linear slopes in the $P$-$W$ dependencies are not unitless, so they cannot be compared to one. In fact, according to the data of \citet[][their Fig.~5b]{dong2019}, rainfall increases faster than moisture content and not vice versa: e.g., while moisture content increases by 1.5-fold from $20$~mm to $30$~mm, rainfall increases two-fold from about $2$~mm/day to $4$~mm/day. A related notice is that precipitation efficiency that \citet{ye2014} define as monthly precipitation divided by mean monthly moisture content and measure it in per cent, is not unitless but has the units of inverse time.)

\begin{figure*}[!tb]
\begin{minipage}{\textwidth}
\centering\includegraphics[width=0.9\textwidth,angle=0,clip]{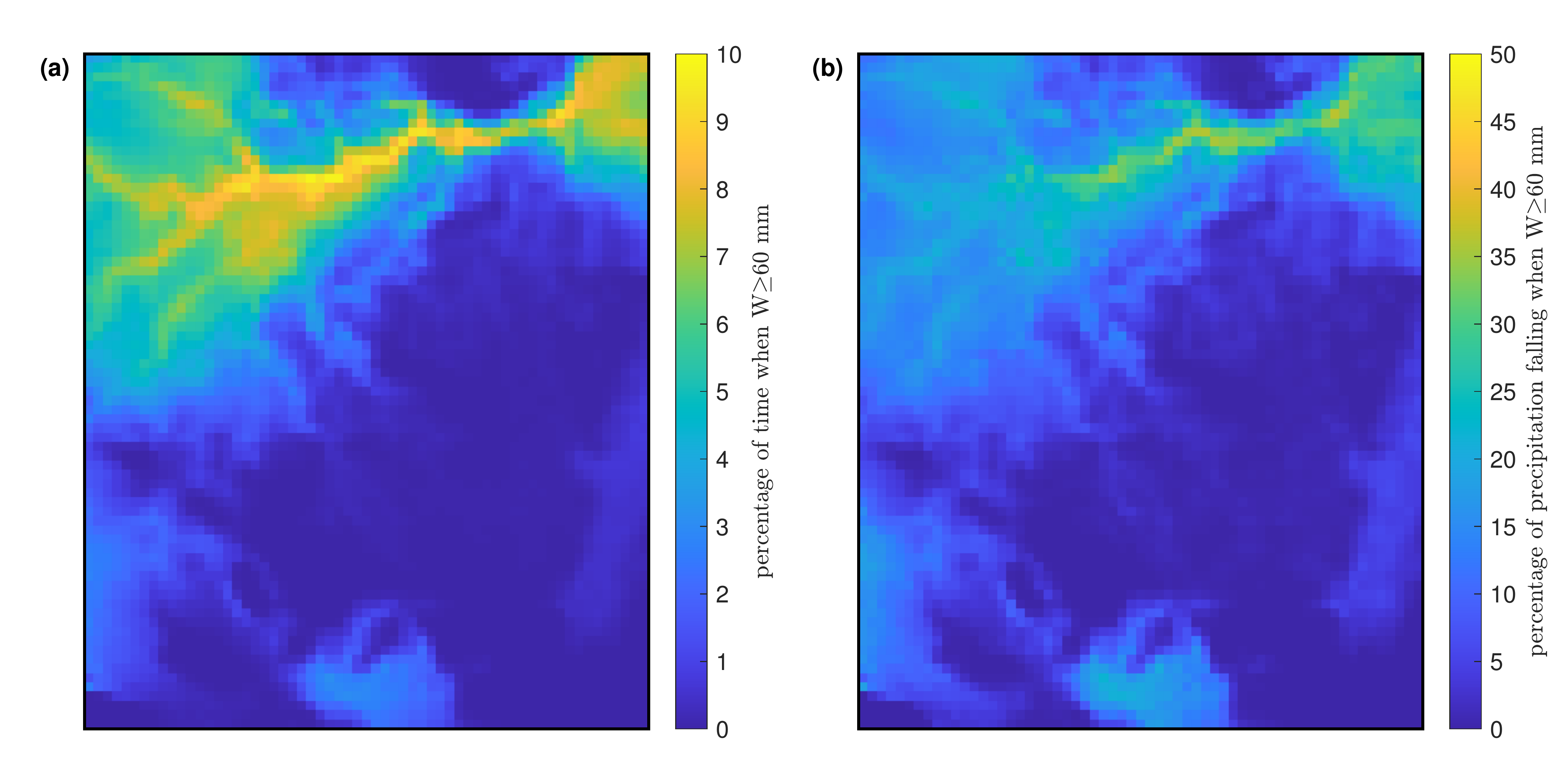}
\end{minipage}%
\caption{
Spatial distribution of the percentage of observations (a) and of total rainfall (b) with $W \ge 60$~mm for the Amazon region studied by \citet{baudena21} (Fig.~\ref{figmap}).
}
\label{highW}
\end{figure*}

The highest values of hourly precipitation and $k_P > 10$ in the Amazon at $W \ge 60$~mm correspond to the opposite case: a saturated atmosphere with ongoing precipitation. Observations with $W \ge 60$~mm constitute $2.4\%$ of all observations and account for $9.3\%$ of all precipitation (Fig.~\ref{highW}).
Among the observations with $W \ge 60$~mm, there are very few ($\sim 3\%$) observations with zero rain. One can hypothesize that these highest values of atmospheric moisture appear as a {\it consequence} of extreme rainfall. The positive dependence between $P$ and $W$ can be explained by the fact that a more intense rainfall is associated with higher convection; hence, a greater part of the atmospheric column is brought close to saturation.

More than $90\%$ of rainfall correspond to the intermediate interval $30~{\rm mm}< W < 60$~mm, which is characterized by a relatively constant $k_P$ (Fig.~\ref{fig3}b). For each $1\%$ of increasing $W$, precipitation rises by $k_P \simeq 6-9$\%.
This indicates that a higher value of $W$ increases the probability of subsequent rainfall (rather than results from rainfall,
as in the case of the highest $W$). 

On a different temporal scale, this dichotomy in the  $P$-$W$ cause-effect relationship was noted for the North Atlantic hurricanes \citep{ar17}. These hurricanes tend to occur where atmospheric moisture content in the preceding days is about $1$~mm higher than the local climatology in the hurricane absence ($43$~mm versus
$42$~mm). When a hurricane does occur, the maximum moisture content is raised by vigorous convection in the windwall up to over $60$~mm \citep[][their Fig.~4d,e]{ar17}. Thus, for intermediate moisture content values, higher moisture content on average causes more precipitation. The highest moisture content values are caused by intense precipitation. The change in $k_P$ behavior at high $W$ may reflect these patterns.

\subsection{Estimating $k_E$}

With known $k_P$ \eqref{P}, we also need to know $k_E$ \eqref{E}, to be able to assess how atmospheric moisture convergence
changes with increasing evapotranspiration from Eq.~\eqref{wbc}.

To find a characteristic value for $k_E$ for the case of a hypothetical deforestation in the Amazon,  we used the following data for the Amazon basin. With mean transpiration $T = 45$~mm/month  \citep{staal2018} and mean annual rainfall of $P =2200$~mm/year \citep{ma06}, the relative reduction of evapotranspiration due to the loss of transpiration from the entire basin is equal to $\Delta E/P = -T/P = -0.25$, i.e., a decrease of 25\% (in proportion to precipitation).  \citet{baudena21} assumed that zeroing transpiration in a region reduces $W$ in a given location by the fraction of water vapor originating from transpiration (i.e., by the recycled moisture content $W_r \le W$) (see Assumption B in Fig.~\ref{conc}). In the Amazon basin, the fraction of water vapor originating from the Amazon transpiration is equal to $W_r/W=0.32$ \citep{staal2018}. Thus, with $\Delta W/W = -W_r/W = -0.32$ and $\Delta E/P = -T/P = -0.25$, we have $k_E = 0.8$.

\begin{figure*}[!bt]
\begin{minipage}[p]{1\textwidth}
\centering\includegraphics[width=0.5\textwidth,angle=0,clip]{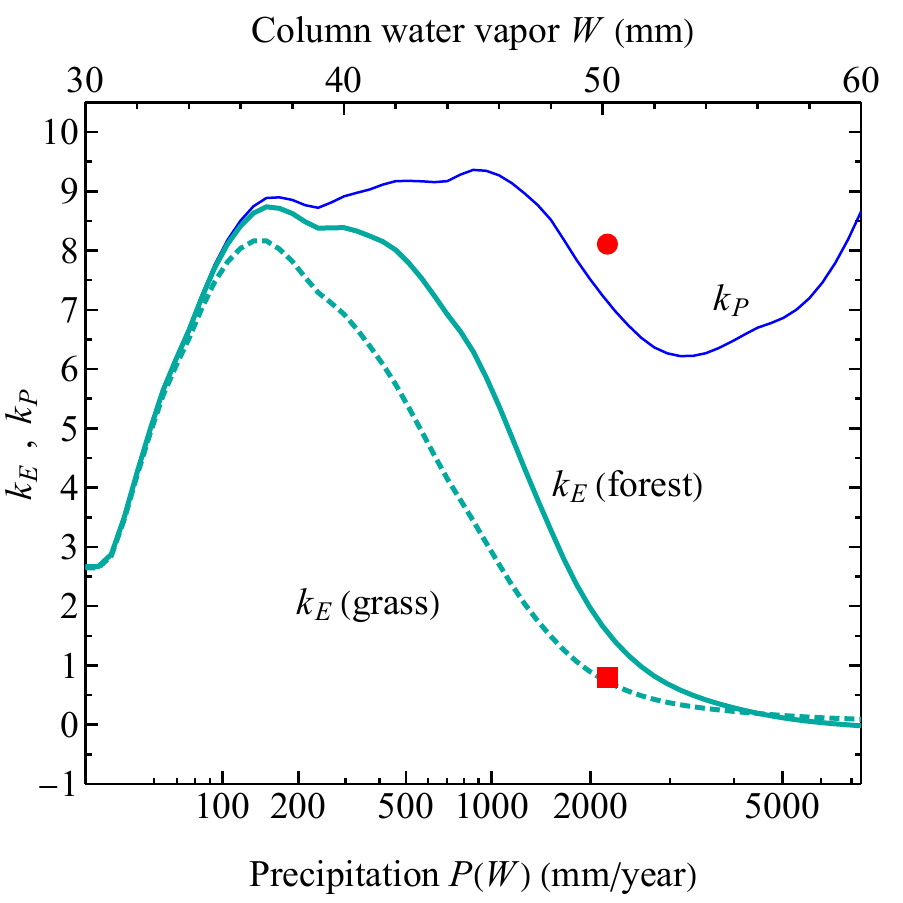}
\end{minipage}
\caption{
Coefficient $k_E$, Eq.~\eqref{E}, determined from Eq.~\eqref{kPE} using $dE/dP$ calculated from Eq.~\eqref{zh} for forests (thick solid curve) and grasses (thick dashed curve), cf. Fig.~\ref{fig2}b. The red square shows $k_E$ estimated independently, assuming that zeroing transpiration in the Amazon ($P = 2200$~mm/year) reduces atmospheric moisture content $W$ by the amount of recycled transpired moisture \citep{baudena21}. 
The red circle shows $k_E$  estimated independently, assuming that zeroing transpiration reduces precipitation $P$ by the flux of recycled transpired moisture (the precipitation
recycling approach, \citep{staal2018}). 
Coefficient $k_P$ is shown for reference (thin blue curve), cf. Fig.~\ref{fig3}b.
}
\label{fig4}
\end{figure*}

To assess the plausibility of this estimate (and the underlying assumption), we combine Eqs.~\eqref{P} and \eqref{E} to find
a relationship between $k_E$ and $k_P$:
\beq\label{kPE}
k_E = k_P \frac{dE}{dP}.
\eeq
Using Eq.~\eqref{zh} to calculate $dE/dP$, we can find $k_E(P)$ from Eq.~\eqref{kPE}. Figure~\ref{fig4} shows that for grasses, we have $k_E = 0.7$ and for forests, we have $k_E = 1.6$ corresponding to the Amazonian rainfall. These estimates provide independent support to $k_E \simeq 1$  that derives from the analysis of \citet{baudena21}.
We recognize the caveat that the data of \citet{zhang2001} used in Fig.~\ref{fig4} are obtained for annual precipitation and evapotranspiration, while $k_P$ is calculated from hourly precipitation data. However, we can see from Fig.~\ref{fig3}b that $k_P$ calculated from monthly and daily precipitation data of \citet{bretherton04} are close (within a factor of two) to hourly data in the interval of moisture content values where $k_P$ is relatively constant.  This suggests that $dE/dP$ can appear roughly time-scale invariant at least at characteristic precipitation rates. However, we must also acknowledge that these patterns require further study.

As illustrated by Fig.~\ref{conc}, the precipitation recycling approach yields a different result for $k_E$. According to \citet{staal2018}, about one-third of the Amazonian precipitation is due to recycled moisture, and of this, two-thirds are due to recycled transpired moisture. If one assumes (see Assumption A in Fig.~\ref{conc}) that zeroing Amazonian transpiration reduces precipitation by $\Delta P/P = -(1/3)\times (2/3) = -0.22$, i.e., by $22\%$, with $\Delta E/P = -T/P = -0.25$, from Eq.~\eqref{kPE} we have $k_E = k_P\times(0.25/0.22)= 8.3 > k_P$ (Fig.~\ref{fig4}). This value, one order of magnitude larger than $k_E = 0.8$ obtained from considering the change in atmospheric moisture content, is too high as compared to what one derives from the data of \citet{zhang2001}. This is unsurprising since the precipitating recycling assumption predicts a decrease in runoff with growing precipitation, while the data of \citet{zhang2001} shown in Fig.~\ref{fig2} indicate an opposite pattern.

\subsection{Ecosystem{\textquoteright}s two moisture regimes}

\begin{figure*}[!tb]
\begin{minipage}[p]{1\textwidth}
\centering\includegraphics[width=0.9\textwidth,angle=0,clip]{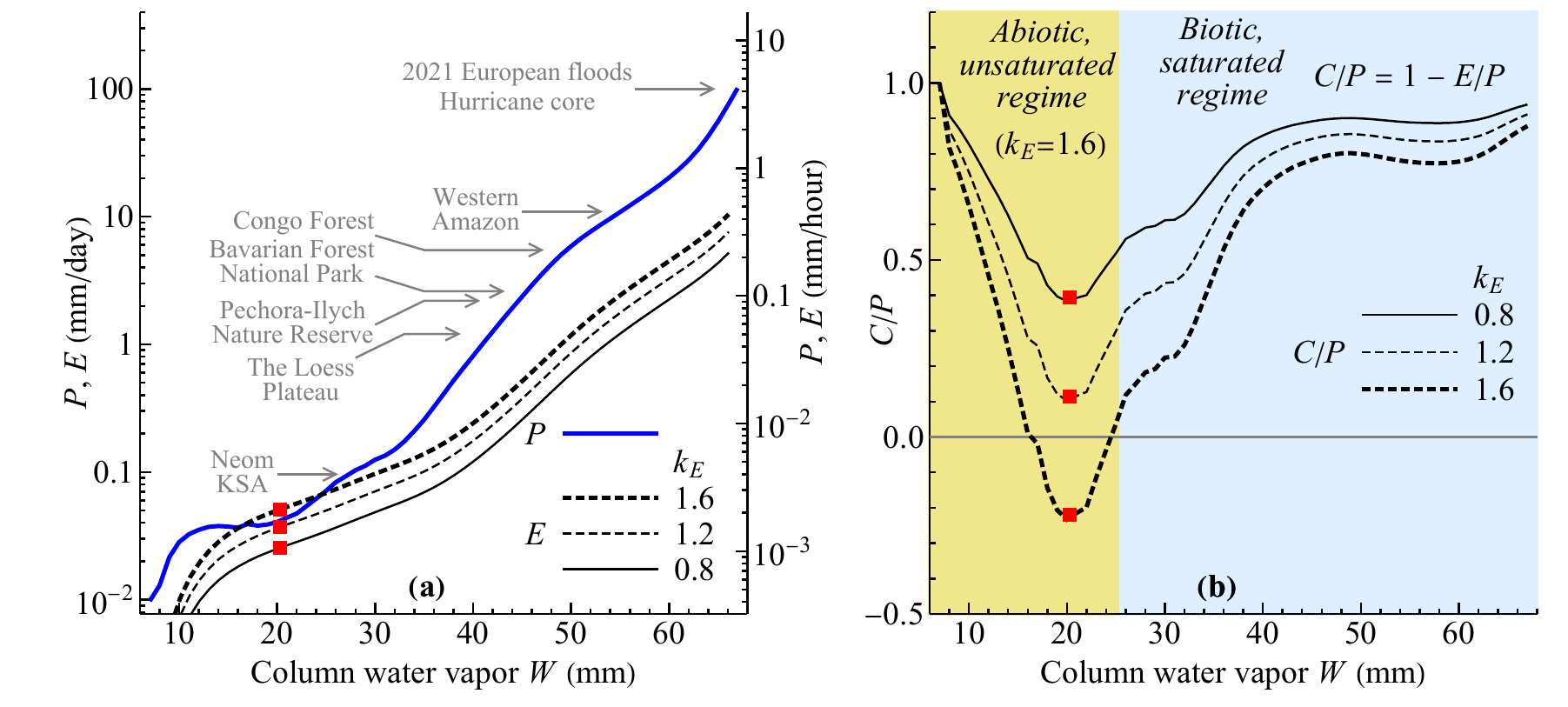}
\end{minipage}
\caption{
Atmospheric moisture budget terms $P$, $E$ and $C$ as related to column water vapor content $W$.  Hourly precipitation data of \citet{baudena21} give the blue curve $P(W)$ in (a),  which we then integrate according to Eq.~\eqref{E} to obtain $E(W)$ (black curves in (a) for different $k_E$). Constrained by the mass balance Eq.~\eqref{wbp} illustrated in Fig.~\ref{fig1}a, we derive moisture convergence $C(W)$ (b) from $P(W)$ and $E(W)$. Arrows indicate characteristic precipitation rates in different locations in the world. Red squares indicate the point where  the moisture convergence (and runoff) are minimal but begin to grow with increasing $E$ at larger $W$.
}
\label{fig5}
\end{figure*}

Using the dependence $P(W)$ (Fig.~\ref{fig3}a) and the estimated value of $k_E$,  we can now integrate Eq.~(\ref{E}) assuming $E(W_{\rm min}) = 0$ for minimal water vapor content $W_{\rm min} = 7$~mm.  This assumption implies that when a steady-state atmosphere is dry, it is because there is no moisture inputs, i.e., the evapotranspiration is negligible.  With known $P(W)$ and $E(W)$ (Fig.~\ref{fig5}a) we find moisture convergence $C(W)$ from Eq.~\eqref{wbc} and evaluate how it depends on $E$.

This reveals two regimes, for low and high $W$ (Fig.~\ref{fig5}b). In the drier regime with $k_P \sim 0$ and $k_P < k_E$, moisture convergence {\it declines} with increasing evapotranspiration and moisture content, while precipitation remains relatively constant.  At higher $k_E$ (i.e., a slower accumulation of atmospheric moisture with growing $E$), there appears an interval of $W$ with negative moisture convergence.  This corresponds to dry conditions when the ecosystem becomes a net source of atmospheric moisture as it transpires at the expense of previously accumulated soil moisture or at the expense of irrigation. We characterize this dry regime as {\textquotedblleft}abiotic{\textquotedblright}, because the ecosystem exploits the geophysical moisture flows and, at $C < 0$, the previously accumulated water stores (Fig.~\ref{fig5}b).

In the wetter regime with $k_P > k_E$, moisture convergence {\it increases} together with evapotranspiration, moisture content, and precipitation. (If the soil moisture content is steady, runoff  $R$ equals moisture convergence $C$ and thus behaves similarly.)
As the air column approaches saturation at high $W$, precipitation begins to increase markedly with $W$ (Fig.~\ref{fig5}a). Evaporation, on the other hand, depends on the moisture deficit near the surface atmosphere, which is largely decoupled from total water vapor content $W$ \citep[][their Fig.~3e,f]{holloway09}. (Indeed, evapotranspiration and moisture convergence have distinct physics. Moisture convergence occurs when the air rises and water vapor condenses.  In contrast, evapotranspiration is not explicitly linked to directional air motions but adds water vapor directly to the atmospheric column facilitated by turbulent diffusion.) Under nearly saturated conditions, evaporation can only proceed if precipitation depletes moisture from the atmosphere creating a water vapor deficit \citep{murakami2006,murakami2021,jimenez2021}. Therefore, at high $W$, precipitation $P$ and evapotranspiration $E$ should  grow approximately proportionally to each other. The evapotranspiration-to-precipitation ratio $E/P$ and the runoff-to-precipitation ratio $R/P = C/P = 1 - E/P$ stabilize at high values of $W$ and then remain approximately constant (Fig.~\ref{fig5}b). In this wet regime, all the components of the water cycle fall under biotic control.

We note that, by using a constant $k_E$ for all values of moisture content, we have obtained
a conservative estimate of moisture convergence at higher moisture content.  This is because with the moisture column approaching
saturation at higher moisture content, increasing evapotranspiration with increasing relative humidity on a larger time scale becomes more difficult. Therefore, on a time scale of days and months (not taking into account intense evaporation during the rainfall), evaporation should increase more slowly with rising moisture content and precipitation (Fig.~\ref{fig2}) causing moisture convergence to raise faster with increasing moisture content than under our conservative assumption of constant $k_E$.

\subsection{Wet season onset in the Southern Amazon}
\label{SA}

Interestingly, relationship \eqref{P} established for local hourly rainfall for a specific study area in the Amazon region \citep{baudena21} encompasses  characteristic precipitation rates over a broad range of spatial and temporal scales, from annual precipitation in deserts \citep{almazroui2020} and semi-arid zones \citep{zhang2022}, boreal \citep{Smirnova2017}, temperate \citep{beudert2018} and tropical \citep{mgl13} forests, to hurricane- \citep{ar17} and flood-causing \citep{kreienkamp2021} rainfall (Fig.~\ref{fig5}a). How universal the dependence of precipitation on atmospheric moisture content could be in different regions and on different time and spatial scales?
What should an experiment look like to demonstrate that ecosystem transpiration can indeed trigger atmospheric moisture convergence on a large
scale? 

First, one would need data on $P$, $E$ and $W$ for a large region, to show that
when evapotranspiration grows, so do atmospheric moisture and precipitation, and that during this process $dP> dE>0$, such
that the atmospheric moisture convergence increases, $dC > 0$. 
Second, one would need data that would prove that the observed increase in moisture content
is indeed caused by enhanced evapotranspiration rather than by an external inflow (i.e., not by an increased $F_i$ in Fig.~\ref{fig1}).
Third, one would need to provide arguments for the cause-effect relationship
to exclude the mere coincidence, i.e., that moisture convergence increased due to external geophysical reasons 
rather than due to local atmospheric moistening by transpiration.

\begin{figure*}[!tb]
\begin{minipage}[p]{1\textwidth}
\centering\includegraphics[width=0.85\textwidth,angle=0,clip]{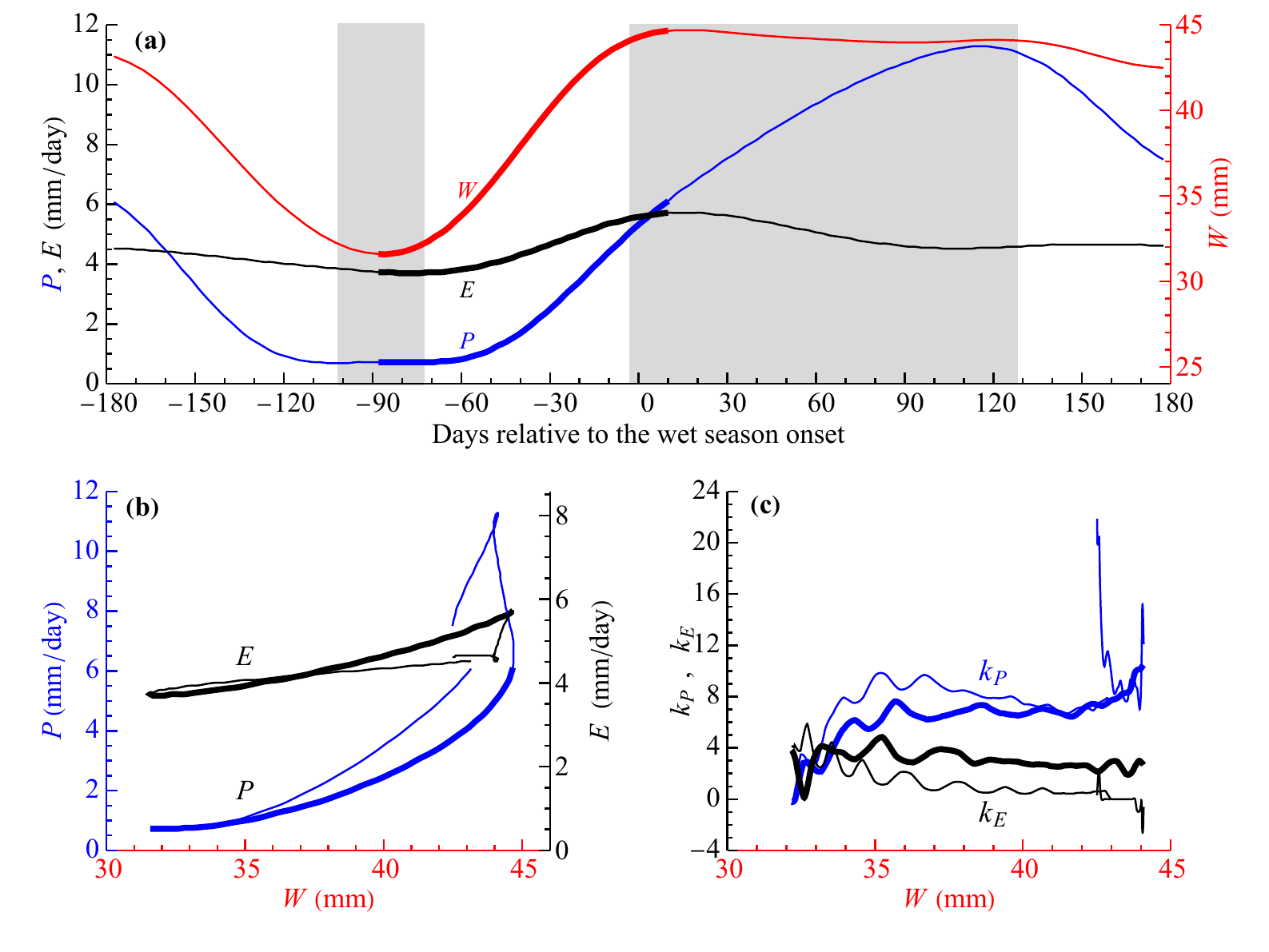}
\end{minipage}
\caption{
The dependence of mean pentad (five days) precipitation $P$ and evaporation $E$ versus columnar water content $W$ 
in the Southern Amazon ($5\degree$S to $15\degree$S, $50\degree$W to $70\degree$W) in $2005$--$2011$ \citep[data in (a) taken from Fig.~S7 of][see Methods]{wright17}. In all panels, solid lines refer to changes from minimum to maximum $W$, and thin lines to the remaining intervals. Gray boxes in (a) indicate intervals with negligible changes of $W$ (defined as $W - W_{\rm min} \le 0.6$~mm and  $W_{\rm max} - W \le 0.6$~mm left and right, respectively). These intervals are excluded in (c), since $dP/dW$ and $dE/dW$ are poorly defined when changes of $W$ are small. 
}
\label{figSA}
\end{figure*}

The onset of the wet season in the Southern Amazon studied by \citet{wright17} meets these requirements. It has long puzzled
researchers as it occurs two months before the major geophysical driver of precipitation at these latitudes:
the seasonal migration of the Intertropical Convergence Zone (ITCZ).
There are no {\it a priori} geophysical grounds to expect the wet season onset so early in this region.

Figure~\ref{figSA}, based on the data of \citet{wright17}, shows that  the annual minimal moisture content coincides with the annual minima of precipitation and evapotranspiration. Prior to the onset of the wet season, the atmospheric moisture content, evapotranspiration, and precipitation all increase. Using data on isotopic fractionation that allow to discriminate water vapor originating from transpiration versus oceanic evaporation, \citet{wright17} concluded that this additional atmospheric moisture comes from evapotranspiration. The established $P(W)$ and $E(W)$ dependencies allow  to calculate $k_P$ and $k_E$ (Fig.~\ref{figSA}b,c). One can see that  $k_P$ is, first, within a similar range $5$--$9$ for most of the $W$ interval as established for the hourly data (cf. Fig.~\ref{fig3}b).  Second, $k_P$ increases with growing $W$, and most conspicuously so at higher $W$ -- this feature is also present in the hourly data. Third, and most important, $k_P > k_E$ (i.e., atmospheric moisture convergence grows with increasing evapotranspiration)  at all $W$ except the lowest and the highest. 

Between the timepoints with minimal and maximum moisture content (thick curves in Fig.~\ref{figSA}), precipitation increases by $157\%$ relative
to the mean precipitation during this period ($\overbar{P} = 3.4$~mm/day = $1240$~mm/year), evapotranspiration increases by $58\%$ relative to this mean precipitation, and moisture content increases by $34\%$ relative to the mean moisture content during this period ($\overbar{W} = 38$~mm). This gives $k_P = 157/34 = 4.6$, $k_E = 58/34 = 1.7$ and $dE/dP = 58/157 = 0.37$. Interestingly, the latter value during the late dry season in the Amazon falls between the corresponding values for grasses ($dE/dP = 0.2$) and forests ($dE/dP = 0.5$) for $P = 1240$~mm/year according to the empirically established dependence of \citet{zhang2001} (Fig.~\ref{fig2}).

At the lowest $W$ with $k_E > k_P \sim 0$ one can suspect the presence of the abiotic regime (Fig.~\ref{fig5}b). 
It can extend with the increasing perturbation of the Amazon forest. At the highest $W$ during the wet season 
(the right gray rectangle in Fig.~\ref{figSA}a) we can see that evapotranspiration declines, but precipitation continues to increase
($k_P$ and $k_E$ are undefined due to $W$ being constant). This looks like an {\textquotedblleft}autocatalytic{\textquotedblright} regime when more rainfall produces more moisture convergence that sustains a high atmospheric moisture content.  Once this regime is switched on, the role of evapotranspiration in the maintenance of moisture content should decline. Studies are ongoing to understand the physics of similar phenomena over the ocean \citep[][]{mapes18,masunaga2020}.

In the view of substantial uncertainties in the observational and model estimates of atmospheric moisture content, moisture convergence 
and runoff, and evapotranspiration \citep[e.g.,][]{allan22,hagemann11,baker21a}, it is essential that $W$ and $C$ 
can be measured independently. Atmospheric moisture content can be assessed
using satellite data like that of the Atmospheric Infrared Sounder (AIRS) \citep{tian19}. Atmospheric moisture convergence $C = R + dG/dt$
can be estimated  from the water balance at the catchment scale from measured streamflow $R$ and estimated groundwater dynamics $dG/dt$ \citep{baker21a}. Using separate $C$ estimates for the Southern and Northern parts of the Amazon basin 
together with the corresponding ERA5 estimates of atmospheric moisture content, we find that their monthly climatologies are similar: 
the minima and maxima fall on the same months (Fig.~S1 in the Supplementary Information). This is consistent
with our proposition that a higher atmospheric moisture content can increase atmospheric moisture convergence. 

Evapotranspiration, estimated as the difference between precipitation and moisture convergence, accumulates uncertainties from both $P$ and $C$ estimates and is the least certain term of the catchment water budget \citep{baker21a}, see also Fig.~S1b.  However, the available data robustly indicate that the climatological month-to-month variability of evapotranspiration (defined as the standard deviation of climatological monthly values, $\sigma_E$) is  small relative to precipitation variability $\sigma_P$, but comparable to that of the atmospheric moisture content, $\sigma_W$.
For the time period $2003$--$2013$ (Fig.~S1), we have $\sigma_E/P = 0.11$, $\sigma_W/W = 0.09$  and $\sigma_P/P = 0.32$ for the Amazon basin as a whole (approximated by the \'{O}bidos gauge), while the corresponding figures for the Southern and Northern parts of the Amazon basin are, respectively, $0.19$, $0.15$, $0.68$ and $0.04$, $0.07$, $0.28$  (see the Supplementary Information).  This is consistent with $k_E/ k_P \sim \sigma_E/\sigma_P < 1$ and $k_P \sim (\sigma_P/\sigma_W) (W/P) \gg 1$. The peculiarities of how evapotranspiration regulates
moisture content require further studies.

\subsection{Ecological restoration at low and high atmospheric moisture: the Loess Plateau example}
\label{LP}

The Loess Plateau in China experienced widespread greening since $1982$. The greening rate has nearly doubled since $1999$ with the implementation
of a state-supported Green for Grain Program (GFGP) \citep{zhang2022}.  The response of the regional hydrological cycle to re-greening has been complex, inconsistent and debated \citep[][and references therein]{feng2016,jia2017,wangXuhu2018,zheng2021,zhang2022,wei2022}.

The Loess Plateau harbors a gradient of summer atmospheric moisture, with the northwest being drier (at $10$--$20$ mm) and the southeast 
wetter ($W$ around 40 mm) \citep{jiang13,dong2019}. These values span abiotic and biotic regimes in Fig.~\ref{fig5}b.
Increased greenness has been associated with a rise of atmospheric water content by about $0.5$--$1$ mm per year across the plateau
\citep{jiang13,tian2022}. Our findings predict distinct responses across the region. Two recent observation-based studies confirm these conclusions
\citep{zheng2021,zhang2022}.

\begin{figure*}[!tb]
\begin{minipage}[p]{1\textwidth}
\centering\includegraphics[width=0.85\textwidth,angle=0,clip]{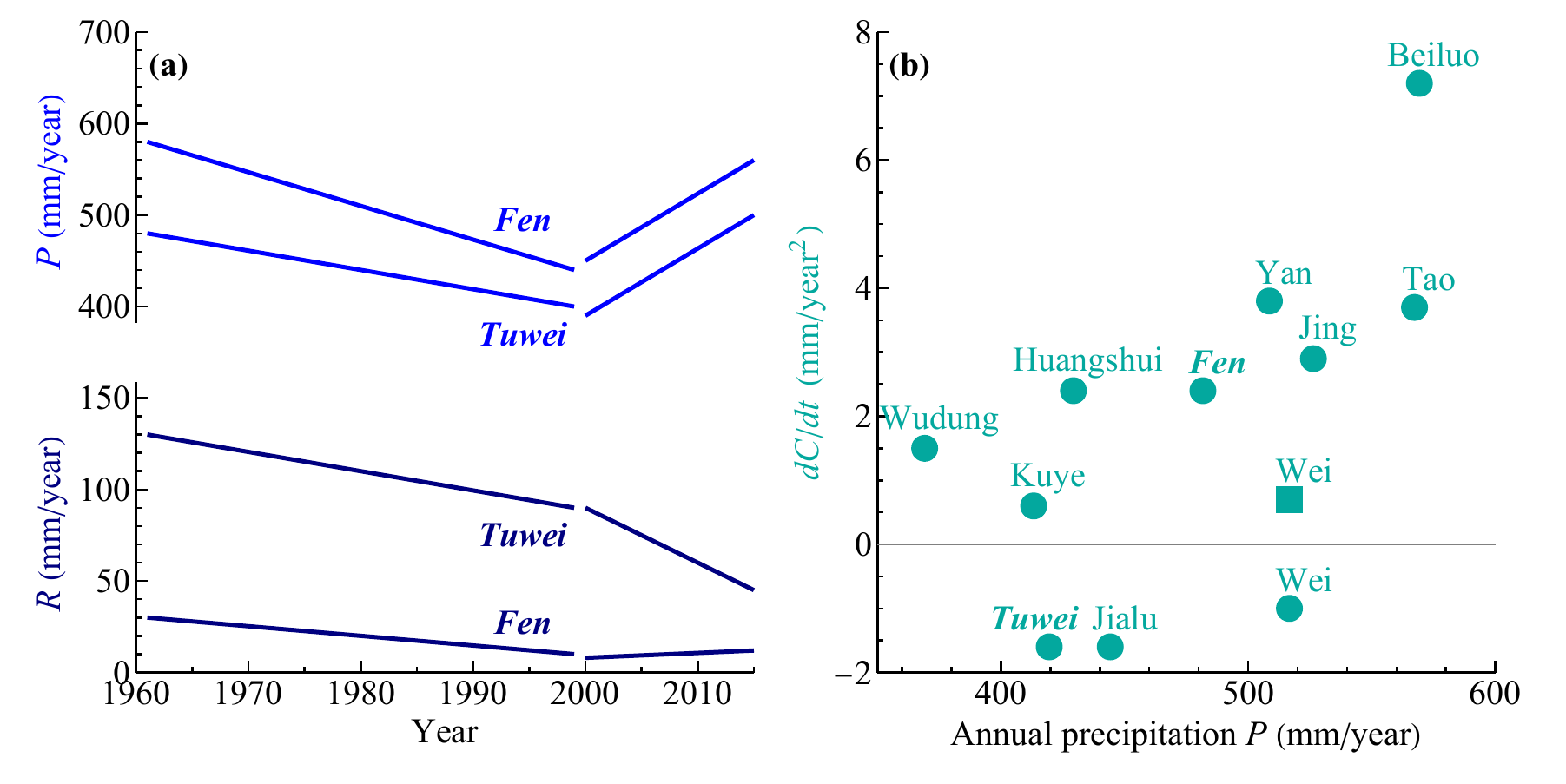}
\end{minipage}
\caption{
Diverse response of the drier and wetter parts of the Loess Plateau to intense re-greening in $1999$--$2015$ according to \citet{zheng2021} (a)
and \citet{zhang2022} (b). In (a), characteristic responses of precipitation and runoff are shown for a drier (Tuwei) and wetter (Fen) river basins;
the data are taken from, respectively, Figs.~5 and~S5 of \citet{zheng2021}. In (b), circles indicate the mean rate of water yield change in $1999$--$2015$ versus the mean annual precipitation in 11 river basins on the Loess Plateau; the data are taken from Fig.~2 of \citet{zhang2022}. The square indicates the mean rate of runoff change in the Wei basin, according to \citet[][their Fig.~S5]{zheng2021}.
}
\label{fig6}
\end{figure*}

According to \citet{zheng2021}, both annual precipitation ($P$) and the runoff-to-precipitation ratio ($R/P$) declined in most river basins on the plateau during $1961$--$1999$, preceding the large-scale re-greening.  When the intense re-greening began in $1999$, precipitation began to increase across the plateau (Fig.~\ref{fig6}a). Meanwhile, the runoff-to-precipitation ratio in the drier northwestern basins continued to decline, while in the wetter southeastern basins (with annual precipitation exceeding $500$ mm/year) it either grew or remained stable, indicating an increasing runoff (Fig.~\ref{fig6}a).
\citet{zheng2021} also noted that soil moisture in the upper 5~cm has increased in the wetter regions in $2000$--$2015$ as compared to $1984$--$1999$,
while in the drier regions it decreased.
 
\citet{zhang2022} used precipitation data and model-estimated evapotranspiration to study the dynamics of the water yield,
\beq\label{wy}
\frac{dC}{dt} = \frac{d(P-E)}{dt} = \frac{dR}{dt} + \frac{d^2G}{dt^2}, 
\eeq
during intense re-greening. 
They also found that in the drier parts of the basin with annual precipitation not exceeding $400$~mm/year, the water yield was declining, while in the wetter parts of the plateau it was increasing (Fig.~\ref{fig6}b). It is important to note that the rate of water yield change does not comprise the rate of ground water (soil water) change ($dG/dt$), but rather the rate at which the groundwater accumulation changes ($d^2G/dt^2$). A positive $d^2G/dt^2$ means that the recharge of groundwater accelerates,
or the decline of groundwater slows down. For example, in $2004$--$2014$ on several sites on the Loess Plateau, the soil water content was shown to decline, $dG/dt<0$, during the first five years of afforestation, but then stabilized ($dG/dt \sim 0$) during the next years \citep{jia2017}. For this period on average $d^2G/dt^2 > 0$, i.e., the increasing water yield has led to a slower decline of soil water content despite the same (or increasing) runoff. 

While there are disagreements between the two studies (e.g., some runoff trends, Fig.~\ref{fig6}b), the general patterns support our propositions. The current mean annual precipitation on the Loess Plateau, $440$~mm/year, appears to be the point of this region{\textquoteright}s transition from the abiotic to the biotic regime.

\begin{table}[htbp]
    \caption{Changes in the water cycle and vegetation parameters of the Loess Plateau between $1982$--$1998$ (pre-GFGP) and
$1999$--$2018$ (post-GFGP) calculated from the data of \citet[][their Fig.~10]{tian2022}. The data refer to the wet season (June-September, approximately
70\% of annual precipitation, e.g., $P=270$~mm/year means that every year there is on average $270$~mm precipitation in the four months from June to September), LAI is leaf area index. }\label{tabLP}
    \begin{threeparttable}
    \centering   
        \begin{tabular}{l l l l l}
        \toprule
Quantity &    Pre-GFGP  &   Post-GFGP  &   Difference $\Delta$ &  Change (\%) \\  \midrule
        \ml{$W$ (mm)}   &   \mc{$\hphantom{11}21.4$}  &   \mc{$\hphantom{11}23.3$}  &  \mc{$\hphantom{-35}1.9$} &  \mc{$\hphantom{-1}9$} \\ 
       \ml{$P$ (mm/year)}   &   \mc{$\hphantom{1}270\hphantom{.4}$}  &  \mc{$\hphantom{1}325\hphantom{.4}$}  & \mc{$\hphantom{-3}55\hphantom{.9}$} & \mc{$\hphantom{-}20$}\\      
        \ml{$E$ (mm/year)}    &   \mc{$\hphantom{1}224\hphantom{.4}$}  &  \mc{$\hphantom{1}247\hphantom{.4}$}  & \mc{$\hphantom{-3}23\hphantom{.9}$} & \mc{$\hphantom{-}10$}\\         
        \ml{$P-E$ (mm/year)}    &   \mc{$\hphantom{11}38\hphantom{.4}$}  &  \mc{$\hphantom{11}70\hphantom{.4}$}  & \mc{$\hphantom{-3}32\hphantom{.9}$} & \mc{$\hphantom{-}84$}\\         
        \ml{LAI (\%)}    &   \mc{$\hphantom{11}96\hphantom{.4}$}  &  \mc{$\hphantom{1}123\hphantom{.4}$}  & \mc{$\hphantom{-3}27\hphantom{.9}$} & \mc{$\hphantom{-}28$}\\         
          \ml{$F_i$ (mm/year)}    &   \mc{$1515\hphantom{.4}$}  &  \mc{$1196\hphantom{.4}$}  & \mc{$-319\hphantom{.9}$} & \mc{$-21$}\\  
          \ml{$F_e$ (mm/year)}    &   \mc{$1483\hphantom{.4}$}  &  \mc{$1132\hphantom{.4}$}  & \mc{$-351\hphantom{.9}$} & \mc{$-24$}\\            
         \ml{$C$ (mm/year)}    &   \mc{$\hphantom{14}32\hphantom{.4}$}  &  \mc{$\hphantom{11}64\hphantom{.4}$}  & \mc{$\hphantom{-8}32\hphantom{.9}$} & \mc{$\hphantom{-}100$}\\  
 \bottomrule                               
     \end{tabular} 
\begin{tablenotes}[para,flushleft]
$k_P = (\Delta P/\Delta W)(\overbar{W}/\overbar{P}) = 2.3$; $k_E = (\Delta E/\Delta W)(\overbar{W}/\overbar{P}) =0.96 < k_P$. 
\end{tablenotes}
\end{threeparttable}
\end{table}

The analysis of the MERRA-2 reanalyses dataset by \citet{tian2022} revealed that the long-term intense re-greening on the Loess Plateau was accompanied by intensification of moisture convergence  (Table~\ref{tabLP}). This suggest that the region is undergoing a transition towards the biotic regime. In twenty years, the net import of moisture has increased almost twofold. This increase occurred despite the decline in the gross import of moisture (i.e., flux $F_i$ in Fig.~\ref{fig1}). 
The gross export of moisture $F_e$, however, declined even more, thus the net moisture income increased.
{\color{black} Since the mean moisture content increased during the same period, the declines in $F_i$ and $F_e$ 
were due to the decline in the mean wind speed. \citet{eiras-barca20} in a model study found that, conversely, deforestation in the Amazon
led to increased wind speeds and reduced rainfall.}
This illustrates that a high gross moisture import does not guarantee an intense water cycle.
Unless the incoming moist air participates in the ascending rain-generating dynamics, it may
flow away again without affecting the local water cycle.

Coefficients $k_P$ and $k_E$ corresponding to the long-term re-greening are, respectively, $k_P = 2.3$ and $k_E = 0.96$ (Table~\ref{tabLP})
as compared to $k_P = 4.6$ and $k_E = 1.7$ for the onset of the wet season in the Amazon (Section~\ref{SA}). 
Besides different spatiotemporal scales and geographic location, the difference between the two patterns concerns
the spatially limited re-greening (and, hence, limited evapotranspiration increase) on the Loess Plateau.
During the onset of the wet season, evapotranspiration in the Amazon increases by about $40\%$ (Fig.~\ref{figSA}a).
On the Loess Plateau, evapotranspiration has increased by about $10\%$ (Table~\ref{tabLP}). This
reflects the limited change in the vegetation cover, which increased by only $8\%$ of the total area, from approximately
$26\%$ in $1982$--$1998$ to $32\%$ in $1999$--$2018$ \citep[][their Fig.~3c]{tian2022}. 
An analysis of isotopic data, similar to the one performed by \citet{wright17} for the Amazon, could
help quantify the contribution of increased evapotranspiration to the added atmospheric moisture.

\section{Discussion}

\subsection{Transpiration triggers moisture convergence}
\label{sec4.1}

We combined two facts: 1) high atmospheric moisture induces precipitation and moisture convergence and 2) transpiration increases atmospheric moisture.  We conclude that transpiration can trigger and control moisture convergence under suitable conditions.

Precipitation occurs when moist air rises and cools. Due to continuity, upward air motion also implies a horizontal air motion and moisture import. The fact that precipitation is associated with high atmospheric moisture content has long been known. Related proposals are that high atmospheric moisture content 
is not simply a result of precipitation (when the rising air brings the atmospheric column closer to saturation) but can itself trigger precipitation and convection \citep[][and references therein]{peters2006,Kuo2017} and that high precipitation can trigger and sustain moisture convergence \citep{mapes18}.
If so, any process that sufficiently increases atmospheric moisture content can also trigger moisture convergence. Transpiration is such a process (Fig.~\ref{fig5}b). This echoes the suggestion of \citet{millan2012,millan2014}, made in the context of the Mediterranean region, when he argued that the rainfall must be {\textquotedblleft}cultivated{\textquotedblright} by maintaining a vigorously transpiring vegetation that can import moisture from the ocean.

Work is in progress to obtain a rigorous theoretical understanding of the transition between non-precipitation and precipitation 
\citep[e.g.,][]{masunaga2020,abbott2021}. While theoretical research is ongoing, the projections of numerical models examining hydrological responses to changes in land cover
remain uncertain. The water yields and runoff in the major river basins are poorly reproduced by global climate models \citep{hagemann11},
and modeling soil moisture has likewise been challenging \citep{zhou2021}. Global climate models cannot reproduce key elements of observed
climatologies without considerable tuning and adjustment \citep{Papadimitriou2017,betts2018,Tan2021}. Therefore,
one cannot rely on the models alone to conceptualize vegetation responses and impacts on the water cycle. A search for new governing
principles to describe these impacts is well justified.

By demonstrating the link between atmospheric moisture content and changes in runoff (water yield), our findings emphasize the need to make atmospheric moisture an integral part of hydrological assessments, along with the conventional analyses of the other water budget terms. Current analyses of water yield are often  decoupled from the analyses of atmospheric moisture content, see, e.g., \citet{lipiao18} and \citet{zhang2022} versus \citet{jiang13} and \citet{dong2019} for China and \citet{feng2021} versus \citet{ye2014} for the boreal zone.

We have pointed out that when the vegetation impact on the water cycle is confined to moisture recycling (i.e., when water vapor is considered as a tracer), then moisture convergence grows as evapotranspiration declines, and vice versa. This pattern has been characterized in the literature as the first-order response of the water cycle to changes in evapotranspiration \citep[e.g.,][]{kooperman18,fowler19} and referred to as supported by overwhelming evidence \citep[][]{ricciardi2022}. In contrast
to moisture recycling, which is recognized as being directly affected by vegetation, moisture supply by atmospheric circulation is more commonly attributed to abiotic factors like climate change and variability \citep[e.g.,][and references therein]{zhang2022}.

Conversely, here we have illustrated that under wetter conditions, increased evapotranspiration should enhance moisture convergence by contributing to increased atmospheric moisture
content and rapidly rising precipitation. We have argued that only under drier conditions, increased evapotranspiration can reduce moisture convergence unless the added atmospheric moisture
is sufficient to significantly enhance precipitation. 

Recent global compilations of the trends in the water cycle are consistent with the proposed duality.
\citet{hobeichi22} ranked the world{\textquoteright}s major regions according to their precipitation and found that in three wettest regions there
are statistically significant increases in precipitation, evapotranspiration and runoff in the analyzed time period $1980$--$2012$ (Table~S2).
In contrast, in the drier regions with increasing evapotranspiration, there is no statistically significant trend in runoff. 
In the drier regions with increasing runoff, there is no statistically significant trend in evapotranspiration.
Furthermore, $dP/dE = k_P/k_E$ ratio declines with increasing aridity (Table~S2), which is consistent with a declining $k_P$ at lower $W$ (Fig.~\ref{fig3}b), increasing $k_E$ at lower $P$ (Fig.~\ref{fig4}), or both.

Another global study indicating distinct responses of the water cycle to increased evapotranspiration under wet and dry conditions, is that of \citet{cui22}. The authors analyzed how changes in leaf area index in the upwind precipitationsheds (LAI\_w) impacted local
precipitation and water yields from $2001$ to $2018$. \citet{cui22} explored the abiotic influences  by performing multiple regression of local annual precipitation $P$ versus LAI\_w and sea surface temperature, net radiation and local surface air temperature.  They found that a global rise in LAI is associated with an increase in annual precipitation of $\Delta P_{\rm LAI\_w} = 11.2$~mm/year in $2001$--$2018$, i.e., by $0.62$~mm/year$^{2}$. The LAI-associated increase in evapotranspiration constituted $58\%$ of the precipitation value, i.e., $0.36$~mm/year$^{2}$. The water yield  accordingly increased by $0.26$~mm/year$^{2}$. 

Comparing these figures with the results of \citet{hobeichi22}, who found global mean $dP/dt = 0.8$~mm/year$^{2}$, $dE/dt = 0.3$~mm/year$^{2}$ and runoff $dR/dt = 0.6$~mm/year$^{2}$ in $1980$--$2012$ (\citet{zhang16} obtained similar results), we conclude that a major part of precipitation and moisture convergence increase on land is likely due to re-greening accompanied by increasing evapotranspiration -- a pattern corresponding to case 3 in Fig.~\ref{fig1}b. This pattern, and not the reduction of moisture convergence with growing evapotranspiration (cases 1 and 2), appears to dominate the global response of the terrestrial water cycle to re-greening (Table~\ref{tabG}). (Note that \citet{cui22} incorrectly stated that the global water availability declines at a rate of $1.76$~mm/year$^{2}$. This contradicts the available evidence (Table~\ref{tabG}).)
At the same time, \citet{cui22} noted that in the drier regions re-greening appeared to reduce water yield. While further studies are clearly needed to refine and consolidate these conclusions, the available data are consistent with our propositions.

\begin{table}[!ht]
     \caption{Global trends (mm/year$^{2}$) in the terrestrial water cycle according to different studies}\label{tabG}
    \begin{threeparttable}
    \footnotesize
    \centering    
    \begin{tabular}{llllllll}
\toprule
    Source & Period & $dP/dt$ &  $dE/dt$ & $dR/dt$ & $dR/dt + d^2G/dt^2$ \\
\midrule
   \citet[][Fig.~3]{gedney06}$^{*}$ & $1960$--$1994$	 &	$-0.23$	&	$-${\it 0.68} 	&	$0.45$  & $0.45$ \\ 
   \citet[][Fig.~4c]{zhang16} & $1981$--$2012$	 &	$\hphantom{-}0.85 \pm 0.5$	&	$\hphantom{-}0.54\pm 0.2$ 	&	& {\it 0.3} \\ 
    \citet{hobeichi22} & $1980$--$2012$	 &	$\hphantom{-}0.8$	&	$\hphantom{-}0.3$ 	&	$0.6$  & {\it 0.5} \\ 
    \citet[][Fig.~3f]{cui22}$^{**}$ & $2001$--$2018$	 &	{$\hphantom{-}\mathbf{0.62}$}	&	{$\hphantom{-}\mathbf{0.36}$} 	& 	  & {$\mathbf{0.26}$} \\   
 \bottomrule   
    \end{tabular}
\begin{tablenotes}[para,flushleft]
Numbers in italics were calculated by us from the water balance,  Eq.~\eqref{wy}, using the data reported in the corresponding source. 
\newline
$^{*}$In the first row, it is assumed that $d^2G/dt^2 =0$.
\newline
$^{**}$Global trends associated with re-greening are shown ($\Delta P_{\rm LAI\_w}/\Delta t$, $\Delta E_{\rm LAI}/\Delta t$ and their difference as water yield, $\Delta t = 18$~years). Partial derivative $\pt P/\pt {\rm LAI\_w}$ was obtained from the multiple regression (see text),
$\Delta P_{\rm LAI\_w} \equiv (\pt P/\pt {\rm LAI\_w})\Delta {\rm LAI\_w}$, where $\Delta \rm LAI\_w$ is the LAI\_w increment in $18$ years. For evapotranspiration $E$, a similar procedure was applied with local LAI,  $\Delta E_{\rm LAI} \equiv (\pt E/\pt {\rm LAI})\Delta{\rm LAI}$.
\end{tablenotes}
\end{threeparttable}
\end{table}

\subsection{Future research questions}

A new fundamental question is  the relative evaluation of the effects of plant transpiration on moisture convergence
versus the larger-scale geophysical climatic impacts. Obviously, not all moisture convergence is due to plant transpiration.
As an example, the mean change of the water yield with precipitation established by \citet{zhang2022} for the Loess Plateau as a whole, see Fig.~\ref{fig2}b,
is about two times the values characteristic of grassland ecosystems (if we use the dependencies in Fig.~\ref{fig2} as a space-for-time substitution). This might be an indication that approximately half of the observed response reflects increased plant transpiration in the region with the rest due to external climatic conditions,
e.g. regional warming and, hence, increased carrying capacity of the atmosphere with respect to the water vapor \citep{tian2022}.
{\color{black} This is consistent with the findings of \citet{cui22}: the trend of moisture convergence attributed to re-greening, $0.26$~mm/year$^{2}$,
 is approximately half of the observed long-term moisture convergence trend established by \citet{hobeichi22} (Table~\ref{tabG}).}

Another important question pertains the spatial scale where vegetation effects can become pronounced.
Regarding spatial scale, \citet{peters2006} found that self-organization of convection over the ocean triggered by high moisture content
can happen over a region of $\sim 200$~km in diameter. In case of a transpiring ecosystem, the spatial scale should depend
on the size of the ecosystem. The Amazon forest can thus modify moisture convergence on a near continental scale (Section~\ref{SA}).

Joint consideration of moist atmospheric processes over land (mediated by terrestrial vegetation)
and over the ocean (largely governed by geophysics alone) can illuminate the nature of both. For example, \citet{mapes18}
emphasized the stable coupling of high oceanic precipitation with moisture convergence (advection) and brought up
the problem of an unclear causality: whether advection causes forced ascent and precipitation, or, conversely,
local convection becomes the cause of a large-scale advection? \citet{mapes18} noted that {\textquotedblleft}a finer
temporal discernment{\textquotedblright} is required to answer this question. \citet{wright17} had already responded to this
question in the Amazon case: using isotopic fractionation, they showed that, at first, transpiration 
moistens the atmosphere facilitating local convection, which leads to large-scale moisture advection.

Given the non-linearity of the precipitation dependence on moisture content, one caveat when considering deforestation and ecorestoration
is that the transient values of precipitation and atmospheric moisture content may not be the same but depend whether the ecosystem is degraded
or restored (i.e., hysteresis) \citep{tewierik2021}. 

When restoration begins from dry condition, often the decline of moisture convergence and runoff can be viewed as a virtue. 
Because under dry conditions most rainfall may occur in violent bursts, after which most water runs away as streamflow without 
being able to infiltrate into the soil, thus not becoming available to plants and not maintaining the photosynthesis. 
Besides, such extreme weather events can cause structural damage to the vegetation cover and promote the soil compaction.
Therefore, the first stage of restoration is to reduce the loss of water from runoff \citep[e.g.,][their Fig.~4]{kravvcik2007,tewierik2021}. Once the hydrological cycle has been sufficiently restored and there is an adequate precipitation and more stable runoff
that can be used for human activities, i.e., near the transition, then there appears a competition for moisture convergence between the recovering ecosystem and people.

Indeed, the decline of the water yield in several reforested areas in China has been interpreted as limiting further ecosystem restoration in these areas \citep[][]{feng2016}. However, Fig.~\ref{fig5}b suggests that if re-greening is continued, the ecosystem can tip into a wetter state when further re-greening enhances both rainfall, moisture convergence, and runoff. Indeed, in the wetter areas in China, re-greening did cause an increase in runoff and water yield \citep[e.g.,][]{wang2018,zheng2021,zhang2022}. Establishing key parameters of the two regimes and assessing the potential transition from the drier to the wetter state (during which the recovering ecosystem might require extra water inputs) can inform and guide afforestation and reforestation strategies, including assessing the possibilities for the recovery of ecosystem productivity in the arid regions. 

\subsection{The role of natural forests, native vegetation and ecological succession}

In a seasonal climate with a pronounced dry and wet season, transpiration during the dry season may cause loss of moisture that will not be recovered during the wet season. How much 
water a recovering ecosystem spends for transpiration at each stage of recovery 
will determine whether the system reaches a successful transition to the wetter, sustainable regime. It is noteworthy that most trees are C3-plants that have a higher {\textquotedblleft}water loss{\textquotedblright} per each carbon dioxide molecule fixed, while the C4-plants that use water more sparingly are predominantly herbs such as grasses and sedges. Accordingly, ecological restoration imitating natural succession with the first stages represented by grasses, with trees appearing on a later (i.e., wetter) stage, should have more chances to achieve a stable wet state than an artificial tree planting (of exotic or native tree species) that transpire actively under the dry conditions and thus can deplete soil  water storage \citep[e.g.,][]{bbli2021,yyang2022}.

The wetter the atmosphere, the stronger the biotic control of the water cycle and the more resilient the forest: by slightly changing evapotranspiration, forests should be able to compensate for unfavorable disturbances of the water cycle, e.g., they can adjust to reduced moisture import by enhancing transpiration-induced moisture convergence. In a drier atmosphere with less rainfall, the control of moisture convergence by transpiration is limited, and the forest is more vulnerable to external perturbations. This could explain why the Amazon forest appears to be losing resilience where the rainfall is already relatively low (rather than where rainfall is decreasing) \citep{boulton2022}. {\color{black} Conversely, external forest disturbances (for example, antropogenic fires) can undermine ecological mechanisms responsible for the maintenance of the wetter regime \citep[e.g.,][]{aleinikov19}. External disturbances can also facilitate a potentially irreversible transition of the ecosystem to the drier regime where it will be losing rather than gaining moisture, with complete ecosystem degradation as a possible outcome. This scenario can stand behind the {\textquotedblleft}landscape trap {\textquotedblright} phenomenon, when repeated disturbances (in particular, logging) increase the ecosystem{\textquoteright}s susceptibility to fire preventing its recovery \citep[][]{lindenmayer22}.}

Regulation of moisture convergence could explain how forests buffer precipitation extremes across continents \citep{oconnor2021,deoliveira2021}. While over a broad range of $30~{\rm mm} \lesssim W \lesssim 60$~mm  the value of $k_P$ is relatively constant (the light blue area in Fig.~\ref{fig3}b), it increases sharply at larger $W$. The interval of $W > 60$~mm harbors very high precipitation rates observed under extreme weather conditions (Fig.~\ref{fig5}a). Volatile organic compounds produced by the forest ecosystem can facilitate precipitation at lower moisture content values, thus not allowing extreme precipitation and winds to develop.  It remains to be investigated whether/how forest disturbances influence the probability of $W$ and $P$ extremes.

Our finding that forest transpiration can trigger and control atmospheric moisture convergence corroborates the biotic pump concept \citep{hess07,jhm14}. The Amazon forest transpiration during the late dry season moistens the atmosphere and triggers the wet season and associated ocean-to-land moisture inflow \citep{wright17}. This dry-season transpiration has a phenological and, hence, an evolutionary component that encodes the Amazon dry-season greening \citep{saleska2016}. Enhanced transpiration preceding the wet season was also observed in the forests of Northeast India and the Congo basin \citep{pradhan19,worder2021}. 

There is a complex interplay of physical and biotic factors behind the ecosystem control of moisture convergence. We have discussed that moisture convergence will grow if precipitation increases faster than evapotranspiration with increasing moisture content (Section~\ref{theor}). Physics dictates that the probability of precipitation rises as the atmosphere approaches saturation. In contrast, {\it other things being equal}, evaporation declines as the water vapor deficit diminishes. Therefore, with increasing $\Delta W > 0$, physics alone guarantees that the required condition $\Delta P > \Delta E$ is fulfilled due to the negative $\Delta E < 0$. 
On the other hand, a declining evapotranspiration cannot cause the moisture content to rise. It is here where the biotic processes come into play. For a given water vapor deficit, evapotranspiration can be increased as young actively transpiring leaves replace old leaves during dry season leaf flush, as seen in the Amazon 
\citep{saleska2016,wu2016,albert2018}.  Leaf flushing can make evapotranspiration grow while (and despite) the atmosphere moistens and the water vapor deficit diminishes, as during the late dry season in the Southern Amazon (Fig.~\ref{figSA}). Once leaf flushing ceases and the atmosphere is sufficiently moistened to enter the {\textquotedblleft}autocatalytic {\textquotedblright}  regime when precipitation and high moisture are maintained by moisture
convergence, the evapotranspiration can decline, as it does with the onset of the wet season in the Southern Amazon (Fig.~\ref{figSA}).

For such a scenario to occur, there should be a proper synchronization between the key geophysical and biotic processes. 
We emphasize the importance of when evapotranspiration increases.
The change in evapotranspiration, as governed by the moisture balance equation, impacts how rapidly the rate of moisture content changes (Section~\ref{theor}). If evapotranspiration begins to rise when the atmospheric moisture content is decreasing
due to geophysical factors, this can partially compensate for the decrease but may be insufficient to revert it.
In such a case, moisture convergence cannot grow, and the transpired moisture can be lost from the ecosystem. {\color{black} On the other hand,
compensation of the decline in the atmospheric moisture content can mitigate potentially greater moisture losses. The ecosystem can react both to the deviation of the atmospheric moisture content from some optimal value as well as to the rate of moisture content change, a similar mechanism to what has been proposed for biotic temperature control \citep{leggett20}. } Native communities, over evolutionary time, evolve phenologies optimized for regional geophysical conditions \citep{gauzere2020}, and the ecohydrological links elucidated here could in principle also select for maximizing the efficiency of the biotic control of the water cycle.
When some set of cultivated species, selected on whatever basis, replaces native ecosystems, this synchronization can be disrupted,
and transpiration can contribute to drying rather than moistening. These complicated processes should be further investigated.

If natural forest ecosystems have indeed evolved mechanisms to stabilize and sustain the continental water cycle, their destruction contributes to the destabilization and impoverishment of regional water cycles and climates. This contribution is underestimated \citep{sheil19}. Future studies of vegetation cover impacts on atmospheric moisture flows must emphasize the role of natural forests \citep{zemp17b,sheil2018,makarieva20b,leitefilho21,hua2022}.

\section{Conclusions}

We have shown that deforestation reduces not just precipitation, but also atmospheric moisture convergence. Conversely, increased transpiration from ecological restoration should increase both -- provided it occurs at a sufficiently high atmospheric moisture content.  These conclusions offer good news for global restoration, where reduced global runoff has often been predicted. Our results indicate that restoration can transition through this negative phase into a context when re-greening can profoundly improve water yields and overall availability. Rigorous interdisciplinary scientific planning merging ecology and atmospheric science is required to achieve this.

\section*{Acknowledgments}
We thank Dr. Arie Staal and Dr. David Ellison for their useful comments on early drafts of this work. The work of A.M. Makarieva is funded by the Federal Ministry of Education and Research (BMBF) and the Free State of Bavaria under the Excellence Strategy of the Federal Government and the L{\"a}nder, as well as by the Technical University of Munich -- Institute for Advanced Study. Thanks to BemTeVi Foundation for support to Ruben Molina under the Windrose Research Grant. Scott Saleska acknowledges support from the U.S. National Science Foundation (award \#2106804).

\clearpage

\appendix
\setcounter{section}{0}%
\setcounter{equation}{0}%
\setcounter{subsection}{0}%
\setcounter{table}{0}%
\setcounter{figure}{0}%
\renewcommand{\theequation}{S\arabic{equation}}%
\renewcommand{\thefigure}{S\arabic{figure}}%
\renewcommand{\thesubsection}{S.\arabic{subsection}}%
\renewcommand{\thetable}{S\arabic{table}}%


\section*{Supplementary Information}
\label{suppl}

\footnotesize
\begin{xltabular}{\textwidth}{XXXXXXXXXX}

\caption{This table reports the $i$-th percentiles $q_i$, the mean ($\overbar{P}$) and the standard deviations (sd) of the distribution of hourly precipitation (mm/hour) in the Amazon study area of \citet{baudena21}, between  0--18$\degree$S and 50--65$\degree$W (Fig.~\ref{figmap}). The analyses were performed for the period 2003--2014 at $0.25\degree$ resolution. Hourly precipitation and column water vapor ($W$) are from the ERA5 dataset \citep{hersbach2018}. Precipitation was binned for different values of $W$, every millimeter of $W$. For each bin, we calculated the precipitation percentiles only if there were more than five points in the bin, from $W=3$~mm up to the maximum retained value of $W = 73$~mm.} \label{tab:long} \\

\hline \hline \multicolumn{1}{l}{$W$ bin} & \multicolumn{1}{l}{$q_{1}$} & \multicolumn{1}{l}{$q_{10}$}  & \multicolumn{1}{l}{$q_{25}$} & \multicolumn{1}{l}{$q_{50}$} & \multicolumn{1}{l}{$q_{75}$} & \multicolumn{1}{l}{$q_{90}$} & \multicolumn{1}{l}{$q_{99}$} & \multicolumn{1}{l}{$\overbar{P}$}  & \multicolumn{1}{l}{sd}\\ \hline 
\endfirsthead

\multicolumn{10}{c}%
{\tablename\ \thetable{} -- continued from previous page} \\
\hline \multicolumn{1}{l}{$W$ bin} & \multicolumn{1}{l}{$q_{1}$} & \multicolumn{1}{l}{$q_{10}$}  & \multicolumn{1}{l}{$q_{25}$} & \multicolumn{1}{l}{$q_{50}$} & \multicolumn{1}{l}{$q_{75}$} & \multicolumn{1}{l}{$q_{90}$} & \multicolumn{1}{l}{$q_{99}$} & \multicolumn{1}{l}{$\overbar{P}$}  & \multicolumn{1}{l}{sd}\\ \hline 
\endhead

\hline \multicolumn{10}{r}{{Continued on next page}} \\ \hline
\endfoot

\hline \hline 
\endlastfoot

     	3 & 0.00 & 0.00 & 0.00 & 0.00 & 0.00 & 0.00 & 0.01 & 0.00 & 0.00 \\ 
     	4 & 0.00 & 0.00 & 0.00 & 0.00 & 0.00 & 0.00 & 0.01 & 0.00 & 0.01 \\ 
     	5 & 0.00 & 0.00 & 0.00 & 0.00 & 0.00 & 0.00 & 0.01 & 0.00 & 0.00 \\ 
     	6 & 0.00 & 0.00 & 0.00 & 0.00 & 0.00 & 0.00 & 0.01 & 0.00 & 0.00 \\ 
     	7 & 0.00 & 0.00 & 0.00 & 0.00 & 0.00 & 0.00 & 0.01 & 0.00 & 0.00 \\ 
     	8 & 0.00 & 0.00 & 0.00 & 0.00 & 0.00 & 0.00 & 0.02 & 0.00 & 0.00 \\ 
     	9 & 0.00 & 0.00 & 0.00 & 0.00 & 0.00 & 0.00 & 0.02 & 0.00 & 0.01 \\ 
     	10 & 0.00 & 0.00 & 0.00 & 0.00 & 0.00 & 0.00 & 0.03 & 0.00 & 0.01 \\ 
     	11 & 0.00 & 0.00 & 0.00 & 0.00 & 0.00 & 0.00 & 0.04 & 0.00 & 0.01 \\ 
     	12 & 0.00 & 0.00 & 0.00 & 0.00 & 0.00 & 0.00 & 0.05 & 0.00 & 0.01 \\ 
     	13 & 0.00 & 0.00 & 0.00 & 0.00 & 0.00 & 0.00 & 0.05 & 0.00 & 0.01 \\ 
     	14 & 0.00 & 0.00 & 0.00 & 0.00 & 0.00 & 0.00 & 0.05 & 0.00 & 0.02 \\ 
     	15 & 0.00 & 0.00 & 0.00 & 0.00 & 0.00 & 0.00 & 0.04 & 0.00 & 0.02 \\ 
     	16 & 0.00 & 0.00 & 0.00 & 0.00 & 0.00 & 0.00 & 0.04 & 0.00 & 0.02 \\ 
     	17 & 0.00 & 0.00 & 0.00 & 0.00 & 0.00 & 0.00 & 0.04 & 0.00 & 0.02 \\ 
     	18 & 0.00 & 0.00 & 0.00 & 0.00 & 0.00 & 0.00 & 0.03 & 0.00 & 0.02 \\ 
     	19 & 0.00 & 0.00 & 0.00 & 0.00 & 0.00 & 0.00 & 0.03 & 0.00 & 0.02 \\ 
     	20 & 0.00 & 0.00 & 0.00 & 0.00 & 0.00 & 0.00 & 0.02 & 0.00 & 0.03 \\ 
     	21 & 0.00 & 0.00 & 0.00 & 0.00 & 0.00 & 0.00 & 0.02 & 0.00 & 0.03 \\ 
     	22 & 0.00 & 0.00 & 0.00 & 0.00 & 0.00 & 0.00 & 0.02 & 0.00 & 0.03 \\ 
     	23 & 0.00 & 0.00 & 0.00 & 0.00 & 0.00 & 0.00 & 0.02 & 0.00 & 0.04 \\ 
     	24 & 0.00 & 0.00 & 0.00 & 0.00 & 0.00 & 0.00 & 0.03 & 0.00 & 0.05 \\ 
     	25 & 0.00 & 0.00 & 0.00 & 0.00 & 0.00 & 0.00 & 0.04 & 0.00 & 0.05 \\ 
     	26 & 0.00 & 0.00 & 0.00 & 0.00 & 0.00 & 0.00 & 0.05 & 0.00 & 0.06 \\ 
     	27 & 0.00 & 0.00 & 0.00 & 0.00 & 0.00 & 0.00 & 0.06 & 0.00 & 0.07 \\ 
     	28 & 0.00 & 0.00 & 0.00 & 0.00 & 0.00 & 0.00 & 0.06 & 0.00 & 0.08 \\ 
     	29 & 0.00 & 0.00 & 0.00 & 0.00 & 0.00 & 0.00 & 0.07 & 0.00 & 0.08 \\ 
     	30 & 0.00 & 0.00 & 0.00 & 0.00 & 0.00 & 0.00 & 0.08 & 0.01 & 0.09 \\ 
     	31 & 0.00 & 0.00 & 0.00 & 0.00 & 0.00 & 0.00 & 0.10 & 0.01 & 0.09 \\ 
     	32 & 0.00 & 0.00 & 0.00 & 0.00 & 0.00 & 0.00 & 0.12 & 0.01 & 0.09 \\ 
     	33 & 0.00 & 0.00 & 0.00 & 0.00 & 0.00 & 0.00 & 0.14 & 0.01 & 0.10 \\ 
     	34 & 0.00 & 0.00 & 0.00 & 0.00 & 0.00 & 0.00 & 0.17 & 0.01 & 0.10 \\ 
     	35 & 0.00 & 0.00 & 0.00 & 0.00 & 0.00 & 0.01 & 0.20 & 0.01 & 0.10 \\ 
     	36 & 0.00 & 0.00 & 0.00 & 0.00 & 0.00 & 0.01 & 0.25 & 0.01 & 0.12 \\ 
     	37 & 0.00 & 0.00 & 0.00 & 0.00 & 0.00 & 0.02 & 0.30 & 0.02 & 0.13 \\ 
     	38 & 0.00 & 0.00 & 0.00 & 0.00 & 0.00 & 0.03 & 0.38 & 0.02 & 0.15 \\ 
     	39 & 0.00 & 0.00 & 0.00 & 0.00 & 0.00 & 0.05 & 0.48 & 0.03 & 0.17 \\ 
     	40 & 0.00 & 0.00 & 0.00 & 0.00 & 0.01 & 0.06 & 0.60 & 0.03 & 0.19 \\ 
     	41 & 0.00 & 0.00 & 0.00 & 0.00 & 0.01 & 0.08 & 0.77 & 0.04 & 0.23 \\ 
     	42 & 0.00 & 0.00 & 0.00 & 0.00 & 0.02 & 0.09 & 0.99 & 0.05 & 0.26 \\ 
     	43 & 0.00 & 0.00 & 0.00 & 0.00 & 0.02 & 0.12 & 1.27 & 0.07 & 0.30 \\ 
     	44 & 0.00 & 0.00 & 0.00 & 0.00 & 0.03 & 0.15 & 1.54 & 0.08 & 0.34 \\ 
     	45 & 0.00 & 0.00 & 0.00 & 0.00 & 0.04 & 0.19 & 1.85 & 0.10 & 0.39 \\ 
     	46 & 0.00 & 0.00 & 0.00 & 0.00 & 0.06 & 0.24 & 2.20 & 0.12 & 0.44 \\ 
     	47 & 0.00 & 0.00 & 0.00 & 0.00 & 0.07 & 0.31 & 2.53 & 0.15 & 0.50 \\ 
     	48 & 0.00 & 0.00 & 0.00 & 0.01 & 0.10 & 0.40 & 2.85 & 0.18 & 0.56 \\ 
     	49 & 0.00 & 0.00 & 0.00 & 0.01 & 0.12 & 0.52 & 3.16 & 0.21 & 0.61 \\ 
     	50 & 0.00 & 0.00 & 0.00 & 0.02 & 0.16 & 0.64 & 3.42 & 0.24 & 0.66 \\ 
     	51 & 0.00 & 0.00 & 0.00 & 0.03 & 0.20 & 0.78 & 3.64 & 0.28 & 0.71 \\ 
     	52 & 0.00 & 0.00 & 0.00 & 0.04 & 0.24 & 0.91 & 3.83 & 0.32 & 0.76 \\ 
     	53 & 0.00 & 0.00 & 0.00 & 0.06 & 0.30 & 1.05 & 4.01 & 0.36 & 0.80 \\ 
     	54 & 0.00 & 0.00 & 0.00 & 0.08 & 0.37 & 1.19 & 4.17 & 0.40 & 0.84 \\ 
     	55 & 0.00 & 0.00 & 0.01 & 0.10 & 0.46 & 1.34 & 4.36 & 0.45 & 0.89 \\ 
     	56 & 0.00 & 0.00 & 0.01 & 0.13 & 0.55 & 1.50 & 4.55 & 0.51 & 0.95 \\ 
     	57 & 0.00 & 0.00 & 0.02 & 0.18 & 0.66 & 1.66 & 4.78 & 0.57 & 1.01 \\ 
     	58 & 0.00 & 0.00 & 0.03 & 0.23 & 0.79 & 1.84 & 5.05 & 0.65 & 1.08 \\ 
     	59 & 0.00 & 0.00 & 0.05 & 0.29 & 0.93 & 2.03 & 5.39 & 0.73 & 1.17 \\ 
     	60 & 0.00 & 0.00 & 0.07 & 0.36 & 1.09 & 2.26 & 5.92 & 0.84 & 1.30 \\ 
     	61 & 0.00 & 0.01 & 0.10 & 0.45 & 1.28 & 2.54 & 6.79 & 0.98 & 1.48 \\ 
     	62 & 0.00 & 0.01 & 0.13 & 0.56 & 1.50 & 2.90 & 8.29 & 1.15 & 1.74 \\ 
     	63 & 0.00 & 0.02 & 0.19 & 0.71 & 1.79 & 3.40 & 10.72 & 1.41 & 2.12 \\ 
     	64 & 0.00 & 0.04 & 0.26 & 0.91 & 2.20 & 4.23 & 13.89 & 1.78 & 2.66 \\ 
     	65 & 0.00 & 0.07 & 0.37 & 1.19 & 2.77 & 5.70 & 16.98 & 2.32 & 3.35 \\ 
     	66 & 0.00 & 0.12 & 0.51 & 1.57 & 3.71 & 8.39 & 20.25 & 3.11 & 4.23 \\ 
     	67 & 0.00 & 0.18 & 0.77 & 2.25 & 5.55 & 11.71 & 22.72 & 4.23 & 5.15 \\ 
     	68 & 0.00 & 0.26 & 1.07 & 3.37 & 8.40 & 14.70 & 25.11 & 5.65 & 6.07 \\ 
     	69 & 0.00 & 0.28 & 1.36 & 4.95 & 11.26 & 16.95 & 26.27 & 7.05 & 6.75 \\ 
     	70 & 0.03 & 0.56 & 2.19 & 6.64 & 13.38 & 19.64 & 29.85 & 8.60 & 7.60 \\ 
     	71 & 0.02 & 0.75 & 2.76 & 10.29 & 15.52 & 21.23 & 29.22 & 10.26 & 7.76 \\ 
     	72 & 0.04 & 0.52 & 3.15 & 11.37 & 17.13 & 20.40 & 28.52 & 10.62 & 7.61 \\ 
     	73 & 0.07 & 0.10 & 0.17 & 0.78 & 2.30 & 29.90 & 35.50 & 6.29 & 11.89 \\ 
\end{xltabular}   

\normalsize

\vspace{1cm}

\begin{figure*}[!ht]
\begin{minipage}[p]{0.8\textwidth}
\centering\includegraphics[width=1\textwidth,angle=0,clip]{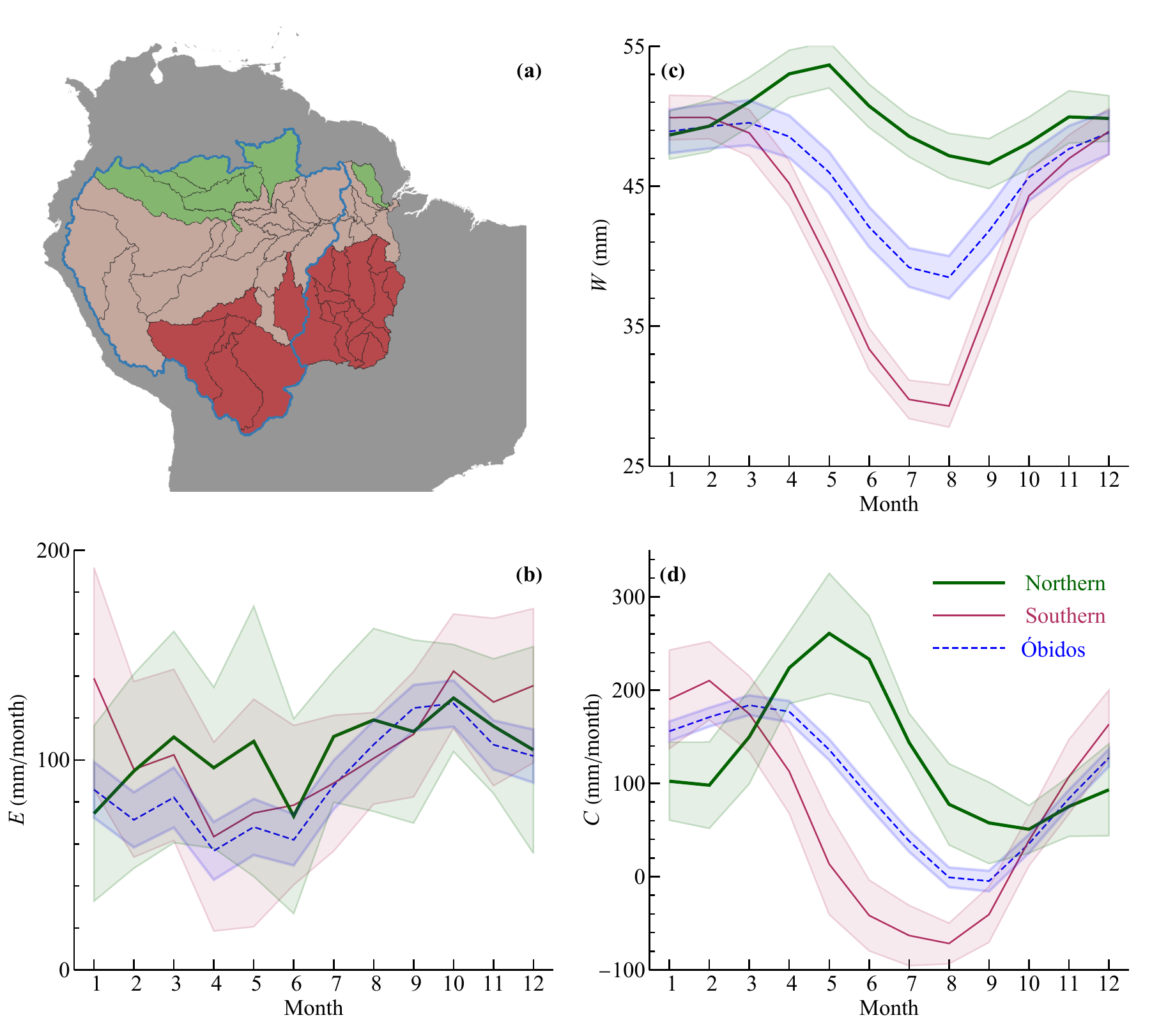}
\end{minipage}

\caption{
Seasonal climatology of evapotranspiration $E$, atmospheric moisture content $W$ and atmospheric moisture convergence $C$ in different parts of the Amazon basin for $2003$--$2013$: {\textquotedblleft}Northern{\textquotedblright} (green color in the map, comprises Japur\'{a}, Negro, Branco, and Jari river basins), {\textquotedblleft}Southern{\textquotedblright} (brownish color in the map, comprises Purus, Madeira, Aripuan\~{a}, Tapaj\'{o}s, and Xingu river basins) and {\textquotedblleft}\'{O}bidos{\textquotedblright} (blue contour in the map, gauged at \'{O}bidos, represents approximately $70\%$ of the Amazon basin), see Table~S1 and Fig.~1 of \citet{baker21a}. Monthly $W$ (assessed as total column water vapor) was computed from the {\textquotedblleft}ensemble\_mean{\textquotedblright} and {\textquotedblleft}ensemble\_spread{\textquotedblright} products in the {\textquotedblleft}reanalysis-era5-single-levels{\textquotedblright} ERA5 dataset (\url{https://doi.org/10.24381/cds.adbb2d47}). First the time-aggregated dataset was calculated (monthly averages of $3$-hourly values), then the basin mask (Southern, Northern, \'{O}bidos) was used to compute spatial averages at each monthly step. Uncertainties represent maximum ensemble spread for the corresponding month in the period $2003$--$2013$. Moisture convergence $C$ calculated from the catchment balance data of \citet{baker21a}: for \'{O}bidos, $C = R + dG/dt$, $R$ and $dG/dt$ are taken, respectively, from 
\citeauthor{baker21a}{\textquoteright}s~\citeyearpar{baker21a} Figs.~S3b and S3c (the uncertainty equals the square root of the sum of squared uncertainties for $R$ and $dG/dt$); for Nothern and Southern parts,  $C = P - E$, where $P$ was retrieved from ERA5 similarly to $W$, while $E$ (assessed from the catchment balance) taken from Fig.~S12 of \citet{baker21a}, uncertainties shown represent the uncertainties in $E$ that in the catchment balance assessment exceed those of $C$ \citep[see][Table~S1]{baker21a}, $E$ for \'{O}bidos, Northern and Southern are, respectively, from Figs.~S3d, S12a and S12b of \citet{baker21a}.
}
\label{figS1}
\end{figure*}

In Fig.~\ref{figS1}, to approximate the Northern and Southern Amazon regions from \citeauthor{baker21a}{\textquoteright}s~\citeyearpar{baker21a}  Figs.~S12, we used basin polygons from HydroBASINS  \citep{lehner13}. For the Amazon basin up to the \'{O}bidos station,  we used the polygon from the Global Runoff Data Base (\url{https://www.bafg.de/GRDC/EN/02_srvcs/22_gslrs/gislayers_node.html}). For geoprocessing, we used geopandas \citep[Ver. 0.12.2;][]{jordahl22} to select and join the HydroBASINS vector data, xarray  \citep[Ver. 2022.12.0;][]{hoyer22} and rioxarray \citep[Ver. 0.13.2;][]{snow22b} to load and subset the ERA5 raster data, and geocube \citep[Ver. 0.3.3;][]{snow22a} to rasterize the regions into the ERA5 data grid. The \'{O}bidos mask has $1534$ cells (the bounding box spans from latitudes $-21$ to $+6$ and longitudes $-80$ to $-55$). The Southern mask has $681$ cells (the bounding box spans from latitudes $-21$ to $-3$ and longitudes $-73$ to $-50$). The Northern mask has $286$ cells (the bounding box spans from latitudes $-4$ to $+6$ and longitudes $-77$ to $-51$). 
 
The quantities $\sigma_E$, $\sigma_W$ and $\sigma_P$, discussed in Section \ref{SA}, 
represent month-to-month variability:
\beq
\sigma_X^2 = \frac{1}{11}\sum_{i=1}^{12} (X_i - \overbar{X})^2 ,
\eeq
where $X_i$ is the value of $X$ in the $i$-th month as shown in Fig.~\ref{figS1}, $\overbar{X}$ is the mean annual value, $X = 
\left\{E,W,P\right\}$ and $P=C+E$. According to ERA5 data, the mean annual precipitation values $\overbar{P}$ are $2800$~mm/year,  $2050$~mm/year and  $2500$~mm/year for Northern, Southern and \'{O}bidos, respectively. The corresponding mean annual atmospheric moisture content $\overbar{W}$ and evapotranspiration $\overbar{E}$ values are $49.7$~mm, $41.9$~mm and $45.5$~mm and $1080$~mm/year,  $1260$~mm/year and  $1250$~mm/year, respectively.

\begin{table}[!t]
     \caption{Characteristic  precipitation $P$ (mm/year), mean trends
(mm/year$^{2}$)  and their confidence intervals (in brackets) for
precipitation $dP/dt$, evapotranspiration $dE/dt$ and runoff $dR/dt$
in world regions with different precipitation regimes, data taken from
Fig.~2 of \citet{hobeichi22}, time period $1980$--$2012$; $k_P/k_E$
calculated as the ratio of $dP/dt$ and $dE/dt$.}\label{tabW}
    \begin{threeparttable}
    \footnotesize
    \centering
    \begin{tabular}{llllllll}
    \hline \hline
    Regime & $P$ & $dP/dt$ &  $dE/dt$ & $dR/dt$ & $k_P/k_E$ \\
       \hline
   {\textquotedblleft}Extremely wet, low
variability{\textquotedblright} &       $4000$ &        $12.5$
($4.4$--$22$)   &       $0.43$ ($0$--$1$)       &       $10.3$ ($5.8$--$15$) & $29$ \\
   {\textquotedblleft}Very wet, low variability{\textquotedblright}
&       $2900$ &        $6.7$ ($2.2$--$11$)     &       $0.55$ ($0.24$--$0.96$) &       $6.9$
($3.0$--$11$) & $12$ \\
   {\textquotedblleft}Wet, low variability{\textquotedblright}
&       $2200$  &       $3.8$ ($0.49$--$8.5$)   &       $0.6$ ($0.28$--$1.0$)   &       $3.5$
($0.94$--$6.4$) & $6.3$ \\
   {\textquotedblleft}Wet, medium variability{\textquotedblright}
&       $1600$ & NS   &       NS    &       NS  &  \\
   {\textquotedblleft}Mild wet, medium variability{\textquotedblright}
&       $1200$  &       NS    &       $0.3$ ($0.04$--$0.57$)  & NS   &  \\
   {\textquotedblleft}Mild dry, medium variability{\textquotedblright}
&       $860$   &       NS    &       $0.54$ ($0.25$--$0.75$) & NS   &  \\
 {\textquotedblleft}Dry, medium variability{\textquotedblright}
&       $560$   &       $0.68$ ($0.08$--$1.3$)  &       $0.43$ ($0.09$--$0.7$)  &       NS  &
$1.6$  \\
 {\textquotedblleft}Dry, high variability{\textquotedblright} & $325$
        &       $0.45$ ($0.01$--$0.95$) &       NS    &       $0.2$ ($0.04$--$0.34$) &   \\
 {\textquotedblleft}Very dry, very high
variability{\textquotedblright} &       $125$   &       NS    &       NS    &       $0.06$
($0$--$0.12$) &   \\
\hline   \hline
    \end{tabular}
\begin{tablenotes}[para,flushleft]
Abbreviation: NS, no significant trend.
\end{tablenotes}
\end{threeparttable}
\end{table}

\clearpage



\end{document}